\documentclass[twocolumn,english,showpacs,superscriptaddress,prb]{revtex4-2}
\usepackage[colorlinks=true,citecolor=blue,linkcolor=blue,urlcolor=blue]{hyperref}
\usepackage[utf8]{inputenc}\usepackage{color}
\usepackage{float}\usepackage{textcomp}
\usepackage{amstext}\usepackage{graphicx}
\usepackage{gensymb}\usepackage{amssymb}
\makeatletter
\usepackage{graphics}\usepackage{subfigure}
\usepackage{longtable}\usepackage{pstricks}
\usepackage{dcolumn}\usepackage{bm}
\usepackage{amsmath}
\usepackage{wasysym}  

\begin{document}
 \title{From spin liquid to magnetic ordering in the anisotropic kagome Y-Kapellasite Y$_3$Cu$_9$(OH)$_{19}$Cl$_8$: a single crystal study}
\author{Dipranjan Chatterjee}
\affiliation{Universit\'{e}  Paris-Saclay,  CNRS,  Laboratoire  de  Physique  des  Solides,  91405,  Orsay,  France}
\author{Pascal Puphal}
\affiliation{Max-Planck-Institute for Solid State Research, Heisenbergstra{\ss}e 1, 70569 Stuttgart, Germany}
\author{Quentin Barth\'elemy}
\affiliation{Universit\'{e}  Paris-Saclay,  CNRS,  Laboratoire  de  Physique  des  Solides,  91405,  Orsay,  France}
\author{Jannis Willwater}
\affiliation{Institut f\"ur Physik der Kondensierten Materie, Technische Universit\"at Braunschweig, Braunschweig, Germany}
\author{Stefan S\"ullow}
\affiliation{Institut f\"ur Physik der Kondensierten Materie, Technische Universit\"at Braunschweig, Braunschweig, Germany}
\author{Christopher Baines}
\affiliation{Laboratory for Muon-Spin Spectroscopy, Paul Scherrer Institut, 5232 Villigen, Switzerland}
\author{Sylvain Petit}
\affiliation{LLB, CEA, CNRS, Université Paris-Saclay, CEA Saclay, Gif-sur-Yvette, France}
\author{Eric Ressouche}
\affiliation{Univ. Grenoble Alpes, CEA, IRIG, MEM, MDN,  38000 Grenoble, France}
\author{Jacques Ollivier}
\affiliation{Institut Laue-Langevin, 38042 Grenoble, France}
\author{Katharina M. Zoch} \affiliation{Physikalisches Institut, Goethe-Universit\"at Frankfurt, Frankfurt am Main, Germany}
\author{Cornelius Krellner} \affiliation{Physikalisches Institut, Goethe-Universit\"at Frankfurt, Frankfurt am Main, Germany}
\author{Michael Parzer} \affiliation{Institute of Solid State Physics, TU Wien, 1040 Vienna, Austria}
\author{Alexander Riss} \affiliation{Institute of Solid State Physics, TU Wien, 1040 Vienna, Austria}
\author{Fabian Garmroudi} \affiliation{Institute of Solid State Physics, TU Wien, 1040 Vienna, Austria}
\author{Andrej Pustogow} \affiliation{Institute of Solid State Physics, TU Wien, 1040 Vienna, Austria}
\author{Philippe Mendels}
\affiliation{Universit\'{e}  Paris-Saclay,  CNRS,  Laboratoire  de  Physique  des  Solides,  91405,  Orsay,  France}
\author{Edwin Kermarrec}
\affiliation{Universit\'{e}  Paris-Saclay,  CNRS,  Laboratoire  de  Physique  des  Solides,  91405,  Orsay,  France}
\author{Fabrice Bert}
\affiliation{Universit\'{e}  Paris-Saclay,  CNRS,  Laboratoire  de  Physique  des  Solides,  91405,  Orsay,  France}

\date{\today}

\begin{abstract}

Y$_3$Cu$_9$(OH)$_{19}$Cl$_{8}$ realizes an original anisotropic kagome model hosting a rich magnetic phase diagram~[M. Hering \emph{et al}, npj Computational Materials 8, 1 (2022)]. We present an improved synthesis of large phase-pure single crystals via an external gradient method. These crystals were investigated in details by susceptibility, specific heat, thermal expansion, neutron scattering and local $\mu$SR and NMR techniques. At variance with polycristalline samples, the study of single crystals gives evidence for subtle structural instabilities at 33~K and 13~K which preserve the global symmetry of the system and thus the magnetic model. At 2.1~K the compound shows a magnetic transition to a coplanar (1/3,1/3) long range order as predicted theoretically. However our analysis of the spin wave excitations yields magnetic interactions which locate the compound closer to the phase boundary to a classical \emph{jammed} spin liquid phase. Enhanced quantum fluctuations at this boundary may be responsible for the strongly reduced ordered moment of the Cu$^{2+}$, estimated to be $\sim$ 0.075~$\mu_B$ from $\mu$SR.

\end{abstract}

\maketitle

\section{Introduction}

Low-dimensional materials with strong magnetic frustration, such as compounds with decoupled antiferromagnetic kagome layers, are prototypical systems to search for an experimental realization of the quantum spin-liquid (QSL) state~\cite{Balents10,Chamorro2020}. This long sought state features no static magnetic order, despite sizeable magnetic interactions, but rather macroscopic entanglement and fractional excitations. Experimental signatures of a QSL state in kagome materials were found first in herbertsmithite, ZnCu$_3$(OH)$_6$Cl$_2$ \cite{mendels07,Han2012} and in the closely related Zn-doped barlowite~\cite{Tustain20,Fu21}. Thus Cu-based systems with a perfect kagome layer have been strongly investigated. 

Around the A$^{2+}$Cu$_3$(OH)$_6$Cl$_2$ family \cite{Puphal2018} a new similar series of compounds has been found with A$^{3+}$Cu$_3$(OH)$_6$Cl$_3$~\cite{Puphal2018,Sun2016,Fu2021}, which realizes perfect kagome layers in a kapellasite-type structure. However, for Y$^{3+}$ the compound is not stable in water solutions, leading to a Cl-OH substitution YCu$_3$(OH)$_{6+x}$Cl$_{3-x}$ with $x=1/3$ (or Y$_3$Cu$_9$(OH)$_{19}$Cl$_{8}$)~\cite{Puphal2017,Barthelemy2019}. This substitution induces a distortion of the magnetic lattice as the Y atoms are moved out of the kagome plane \cite{Puphal2017}. The distorted kagome lattice of Cu$^{2+}$ ions is scarcely studied with the exception of the family around (Cs,Rb)$_2$Cu$_3$(Ti,Sn,Zr,Hf)F$_{12}$ \cite{Mueller1995,Matan2019,Downie2015,Grbi13}. However recent structural investigations of low temperature diffraction on various kagome systems show that a structural instability induces distortions in barlowite, claringbullite~\cite{Henderson2019}, volborthite~\cite{Ishikawa2015} and vesigneite~\cite{Boldrin2016} making it a typical structural motif at low temperatures. 

In Y-kapellasite ($x=1/3$) that we investigate in this article, the distortion yields two nonequivalent Cu sites and a unique magnetic model with three different nearest neighbor interactions, while still retaining rotational symmetry around the hexagons of the anisotropic kagome lattice (see inset Fig.~\ref{fig:1} and Ref.~\cite{Barthelemy2019}). The discovery of this compound has triggered a detailed theoretical study~\cite{hering2022} which disclosed the full classical phase diagram of the model and its richness. Interestingly, besides two long range ordered phases with propagation vectors $Q=(1/3,1/3)$ and $Q=(0,0)$, a large area in the phase diagram, coined "classical spin liquid" phase, which encompasses the isotropic kagome model, opens up. It remains so far largely unexplored, although it could realize an unprecedented "jammed" spin liquid phase, characterized by a discrete ground state degeneracy, in a disorder free model. Besides, first principle calculations confirm the relevance of this anisotropic nearest-neighbor model for Y-kapellasite and locates it in the $(1/3,1/3)$ long range ordered phase~\cite{hering2022}. Here, we report a comprehensive study of the physical properties of large phase pure single crystals that can be compared to the theoretical prediction and which show marked deviations with respect to former studies of polycristalline samples~\cite{Barthelemy2019} or other synthesis route for crystals~\cite{Sun2021}.

The paper is organized as follows. In sections II and III the synthesis of phase-pure large single crystals is described and macroscopic measurements of the susceptibility, heat capacity and thermal expansion are presented, providing evidence for a series of structural and magnetic transitions below 33~K. In sections IV and V, the transitions are investigated in detail by neutron diffraction and by Cl NMR, which shows in particular good agreement with the theoretically predicted spin order in the ground state. In section VI, muon spin spectroscopy is used to prove the bulk nature of the 2.1~K magnetic transition. In section VII, the spin wave excitations are studied by inelastic neutron scattering and give a novel insight into the determination of the magnetic interactions beyond first principle calculations. Finally in section VIII, we summarize and discuss our results in comparison with previous ones on polycristalline samples and in light of theoretical studies.


\section{Synthesis and susceptibility}

The crystal growth of Y-Kapellasite was originally reported in Ref.~\cite{Puphal2017}, where 0.59~g Y$_2$O$_3$, 0.82~g CuO and 0.89~g CuCl$_2\cdot$2(H$_2$O) in 10~ml H$_2$O were heated up to the dissolution point of Y$_2$O$_3$, followed by a slow cooling to crystallize. However, these crystals of an average size of 1x1x1~mm$^3$ suffered small CuO inclusions since the growth takes place on the surface of the polycrystalline CuO starting material, as the dissolution point of Y$_2$O$_3$ and the crystallization point of the compound lie above the maximum in solubility of CuO.
Reference~\cite{Sun2021} describes a synthesis of inclusion free crystals using LiOH, Y(NO$_3$)$_3\cdot6$H$_2$O and CuCl$_2\cdot2$H$_2$O. However the reported crystals are somewhat disordered as a partial occupation of the Y site is observed, similarly to what was found in the related compound YCu$_3$(OH)$_6$Cl$_3$ \cite{Sun2016}. This likely hints at a phase mixture in both cases due to contact with water. Indeed, a low percentage of occupation of Y for YCu$_3$(OH)$_6$Cl$_3$~\cite{Sun2016} lies out of the kagome plane, which represents structural parts of Y-Kapellasite Y$_3$Cu$_9$(OH)$_{19}$Cl$_8$. For the reported inclusion free Y$_3$Cu$_9$(OH)$_{19}$Cl$_8$~\cite{Sun2021}, the partial occupation of Y can be explained similarly, since the specific heat shows a mixture of the signal observed for $x = 0$ and  $x = 1/3$ as shown in Fig.~\ref{Specific heat}.

\begin{figure}[t]
\begin{centering}
\includegraphics[width=\columnwidth]{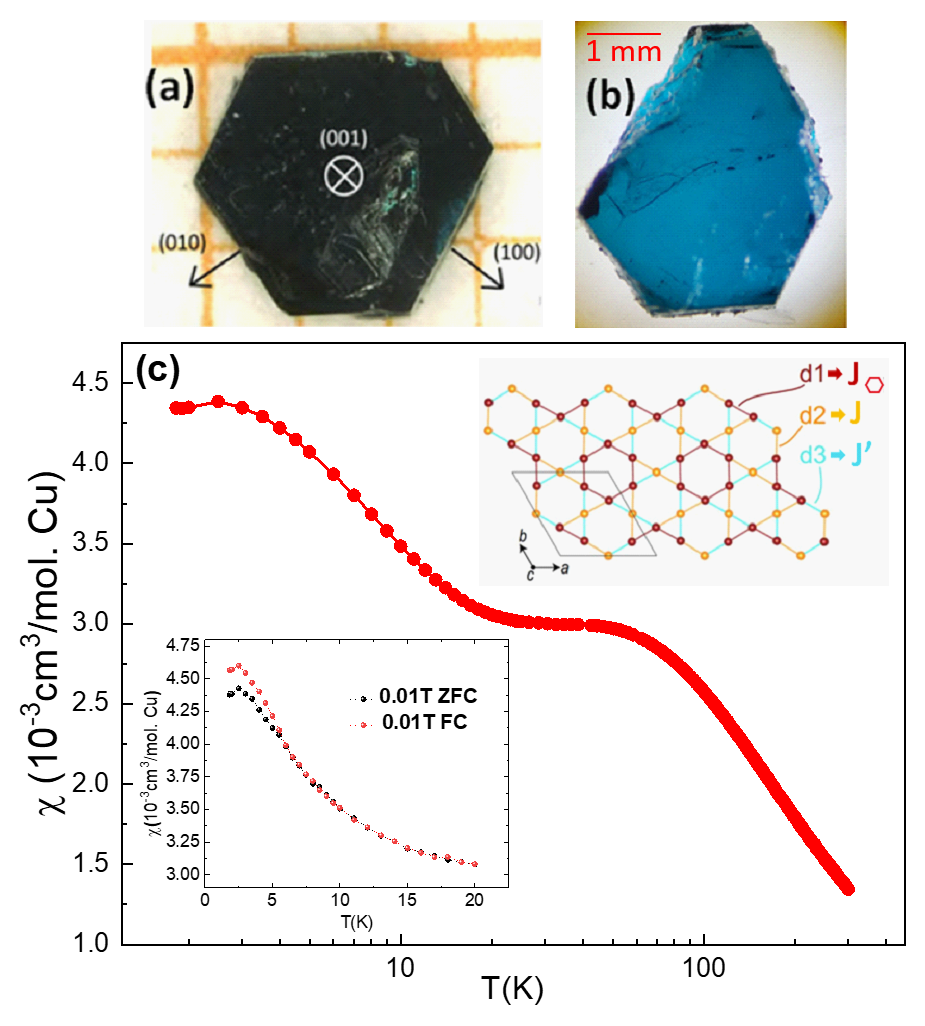}
\par\end{centering}
\caption{a) Y$_3$Cu$_9$(OH)$_{19}$Cl$_8$ single crystal along the $c$ axis. b) Image in transmission light of an as grown single crystal. c) Susceptibility (M/H) versus temperature with a 1~T field applied along $c$. Bottom left inset: field cooled (FC) and zero field cooled (ZFC) measurements at 0.01~T. Upper right inset: anisotropic kagome lattice with 3 different nearest neighbor Cu-Cu bonds and the related 3 different magnetic interactions.
}
\label{growth}
\end{figure}

We thus improved the synthesis for inclusion-free large bulk single crystals suitable for neutron studies via a horizontal external gradient growth method in a thick-walled quartz ampule as described in detail for herbertsmithite in Ref.~\cite{Han2011}. The growth is realized by slowly dissolving CuO in a YCl$_3$-H$_2$O (or YCl$_3$-D$_2$O for deuterated crystals) solution and transporting it to the cold end. The growth is executed in a three zone furnace with a gradient of 25$\degree$C and a temperature of 240$\degree$C at the hot end (note that the elevated pressure at this temperature requires especially thick quartz ampules). The gradient was optimized as too low temperatures yielded a mixture of Y-Kapellasite and Clinoatacamite. Afterwards, the inclusion-free hexagonal single crystals have an average size of 3x3x1~mm$^3$ up to 3x3x3~mm$^3$, if grown over several weeks. The pictures of single crystalline Y-kapellasite are displayed in Fig.~\ref{growth} a) and b), which show a transparent specimen without any visible impurity inclusions.

The dc-susceptibility of a 15.7~mg single crystal of Y-Kapellasite measured in a Quantum Design MPMS XL7 squid system with a 1~T field applied along the $c$ axis is shown in Fig.~\ref{growth}~c).  It reproduces published results \cite{Puphal2017,Biesner2022}, with an antiferromagnetic Curie-Weiss constant $\theta_{\rm{cw}}=102(1)$~K and $g_{\rm{c}}=2.40(1)$. Measurements in a lower 0.01~T magnetic field, shown in the inset, reveal a bifurcation of the field cooled and zero field cooled curves below $\sim 5$~K, and a maximum around 2.5 K. These low field measurements point at a magnetic transition which is investigated in details with neutron scattering and resonance techniques in the following sections. At variance with reference~\cite{Sun2021}, we observe no magnetic transition at 11~K. Such a transition, reminiscent of the one observed at 12~K in the $x=0$ counterpart, supports a phase mixture scenario for the crystal of reference~\cite{Sun2021}.  

\section{Specific heat and thermal expansion}

\begin{figure}[t]
\begin{centering}
\includegraphics[width=1\columnwidth]{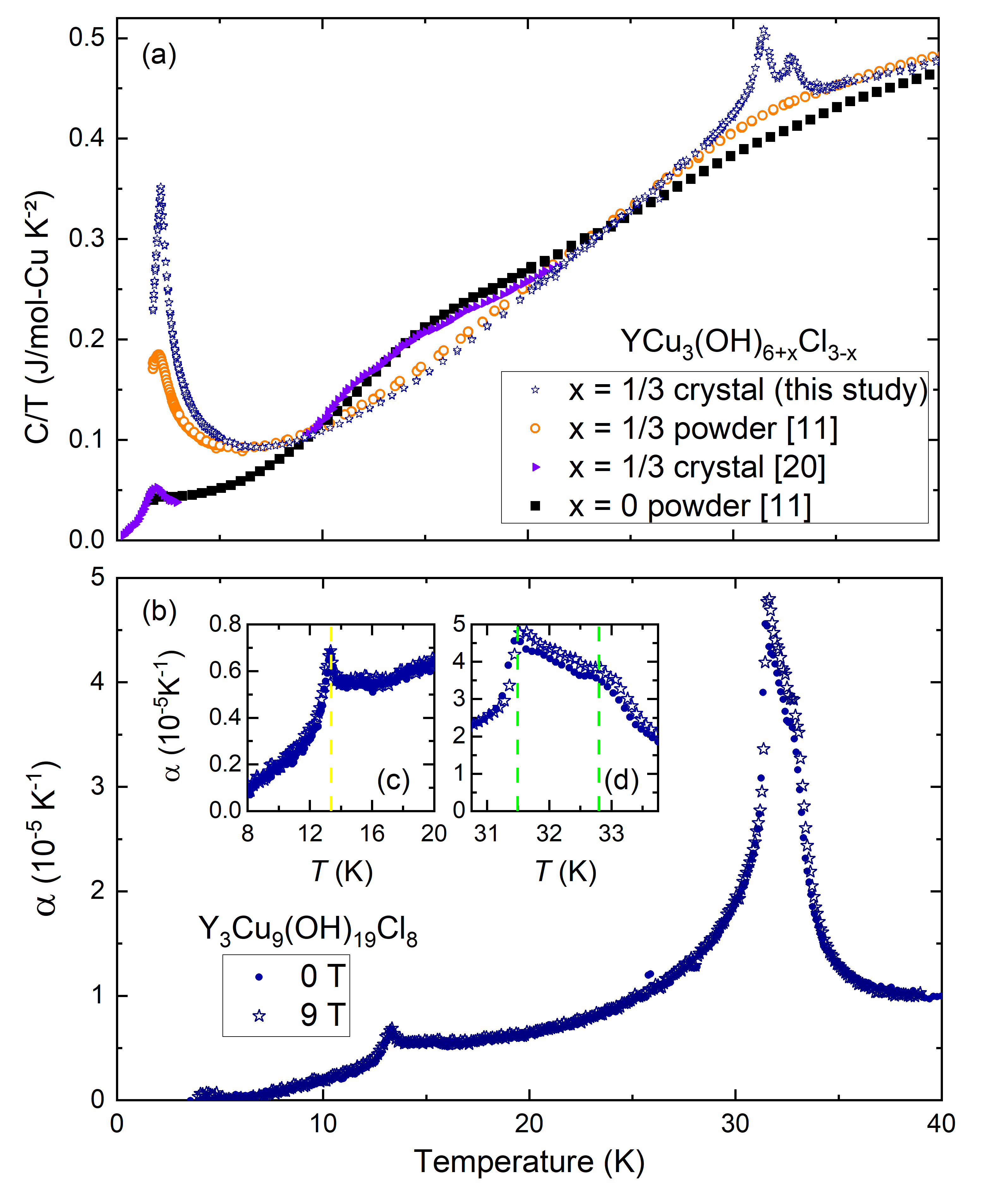}
\par\end{centering}
\caption{a) Specific heat divided by temperature in the temperature range from 1.8 to 40 K for the $x=1/3$ optimized large crystals, compared to literature results for $x=1/3$ and $x=0$ powder samples~\cite{Barthelemy2019} and the reported inclusion free $x=1/3$ crystals from Ref.~\cite{Sun2021}. b) Thermal expansion along the $s$ direction of an optimized large Y-Kapellasite single crystal. c) and d) Insets: enlarged views of the peaks detected in main panel b). Note the different vertical scales in the two insets.}
\label{Specific heat}
\end{figure}

Specific heat was measured with the standard option of a Quantum Design Physical Properties Measurement System (PPMS), using a 5.935~mg inclusion-free single crystal formerly described. Its temperature evolution is shown in Fig.~\ref{Specific heat}~a) and compared to the published results for $x=0$ and $x=1/3$ polycrystalline samples~\cite{Barthelemy2019} and the $x=1/3$ single crystal reported in Ref.~\cite{Sun2021}. Unexpectedly, our $x=1/3$ single crystal presents a clear double peak at 33~K, which is absent in powder samples of the same compound. The underlying entropy amounts to 0.1537~J/mol~K,  below 3\% of $R$ln2, calculated by using the powder sample as a baseline. There is no associated sharp feature in the susceptibility in this temperature range so that the double peak is likely signaling a structural transition. Note that this 33~K anomaly was not reported in the single crystals of Ref.~\cite{Sun2021} and that, on the contrary, the large hump around 15~K seen in the latter crystals and in the $x=0$ powder sample, associated to magnetic transitions, is absent from our phase-pure $x=1/3$ single crystal. Only at 2.1~K do we find a sharp peak in specific heat, which was already reported for this compound both in powder and crystals~\cite{Puphal2017,Barthelemy2019}, and stems from a magnetic long range ordering transition as shown later in this article. For the disordered powder sample $x=1/3$, the underlying entropy of the transition is reduced and previous muon spin relaxation studies had shown the absence of long range order~\cite{Barthelemy2019}, possibly arising from the slightly altered structure due to the suppression of the 33~K transition.

To inspect the 33~K anomaly in more detail, we performed thermal expansion experiments along the crystallographic $c$-axis, i.e. perpendicular to the kagome layers, using a capacitive dilatometer. Figure \ref{Specific heat} b) displays the thermal expansion coefficient $\alpha$ on the same temperature scale as the specific heat data in panel a). We find a similar anomaly below 33~K as in $C/T$, consisting of two slightly broadened, and thus overlapping peaks with a ratio between the sizes of the higher-temperature and lower-temperature peaks comparable to a). The relative length change incorporated in these two sharp features amounts to $\Delta L/L = 10^{-4}$. 
In addition, we discover another, significantly smaller anomaly around 13~K, which can be seen in more details in the inset c). Below 5~K an anomaly in $\alpha$ forms as approaching the transition at 2.1~K, signalling the importance of magneto-elastic coupling as antiferromagnetism sets in. Overall, the anomalies at 13~K and 33~K reveal clear structural effects in this compound with a distorted kagome lattice, without noticeable changes upon applying a magnetic field of 9~T along the $c$-axis.

\section{Diffraction}

\begin{table*}
\scriptsize
\begin{tabular}{l|lll|lll|lll|lll}
T (K) & \multicolumn{3}{l|}{0.065} & \multicolumn{3}{l|}{8} & \multicolumn{3}{l|}{20} & \multicolumn{3}{l}{40}\tabularnewline
\hline 
 & x/a & y/b & z/c & x/a & y/b & z/c & x/a & y/b & z/c & x/a & y/b & z/c\tabularnewline
 \hline 
Y1 & 0 & 0 & 0.1318(7) & 0 & 0 & 0.1314(6) & 0 & 0 & 0.1324(7) & 0 & 0 & 0.1302(5) \tabularnewline
Y2 & 0 & 0 & 0.5 & 0 & 0 & 0.5 & 0 & 0 & 0.5 & 0 & 0 & 0.5 \tabularnewline
Cu1 & 0.6661(10) & 0.8325(11) & 0.5025(4) & 0.6658(10)& 0.8340(11) & 0.5021(4) & 0.6675(11) & 0.8328(12) & 0.5026(4) & 0.6645(7) & 0.8317(7) & 0.5021(3) \tabularnewline
Cu2 & 0.5 & 0 & 0.5 & 0.5 & 0 & 0.5 & 0.5 & 0 & 0.5 & 0.5 & 0 & 0.5 \tabularnewline
Cl1 & 0.6745(9) & 0.0033(10) & 0.2845(2) & 0.6743(8) & 0.0037(10) & 0.28435(19) & 0.6748(9) & 0.0031(11) & 0.2845(2) & 0.6732(6) & 0.0027(6) & 0.28395(15) \tabularnewline
Cl2 & 0 & 0 & 0.3368(5) & 0 & 0 & 0.3368(5) & 0 & 0 & 0.3365(6) & 0 & 0 & 0.3373(4) \tabularnewline
O1 & 0 & 0 & 0 & 0 & 0 & 0 & 0 & 0 & 0 & 0 & 0 & 0 \tabularnewline
D1 & 0.456(9) & 0.776(10) & 0.660(4) & 0.455(9) & 0.783(10) & 0.664(4) & 0.455(9) & 0.776(10) & 0.660(4) & 0.462(7) & 0.777(7) & 0.661(3) \tabularnewline
O2 & 0.8119(14) & 0.8060(12) & 0.5435(7) & 0.8122(13) & 0.8046(12) & 0.5438(7) & 0.8129(14) & 0.8060(13) & 0.5438(8) & 0.8121(9) & 0.8035(9) & 0.5435(4) \tabularnewline
D2 & 0.7769(19) & 0.7778(17) & 0.6011(9) & 0.7874(18) & 0.7858(17) & 0.5991(9) & 0.7723(17) & 0.7722(16) & 0.6030(8) & 0.7866(13) & 0.7827(13) & 0.6005(6) \tabularnewline
O3 & 0.5293(12) & 0.6635(11) & 0.5540(6) & 0.5285(11) & 0.6638(10) & 0.5537(6) & 0.5283(12) & 0.6622(11) & 0.5542(6) & 0.5290(8) & 0.6634(8) & 0.5543(4) \tabularnewline
D3 & 0.5494(17) & 0.6731(18) & 0.6065(8) & 0.5575(16) & 0.6709(17) & 0.6082(8) & 0.5452(15) & 0.6740(18) & 0.6051(8) & 0.5548(12) & 0.6692(13) & 0.6069(5) \tabularnewline
O4 & 0.5142(12) & 0.8433(11) & 0.4631(5) & 0.5167(11) & 0.8430(10) & 0.4632(5) & 0.5154(12) & 0.8450(11) & 0.4634(6) & 0.5162(8) & 0.8441(8) & 0.4638(4) \tabularnewline
D4 & 0.5008(16) & 0.8310(17) & 0.4096(8) & 0.5013(16) & 0.8328(16) & 0.4097(8) & 0.5018(17) & 0.8320(17) & 0.4099(8) & 0.5029(12) & 0.8337(12) & 0.4099(6) \tabularnewline
\label{Rietveld}
\end{tabular}

\caption{Rietveld refinement results of neutron diffraction data obtained at
65 mK, 8 K, 20 K and 40 K with R-3 (\#148) $a=b=11.539$ \AA, and $c=17.1355$ \AA.}
\end{table*}

\begin{figure}[b]
\begin{centering}
\includegraphics[width=1\columnwidth]{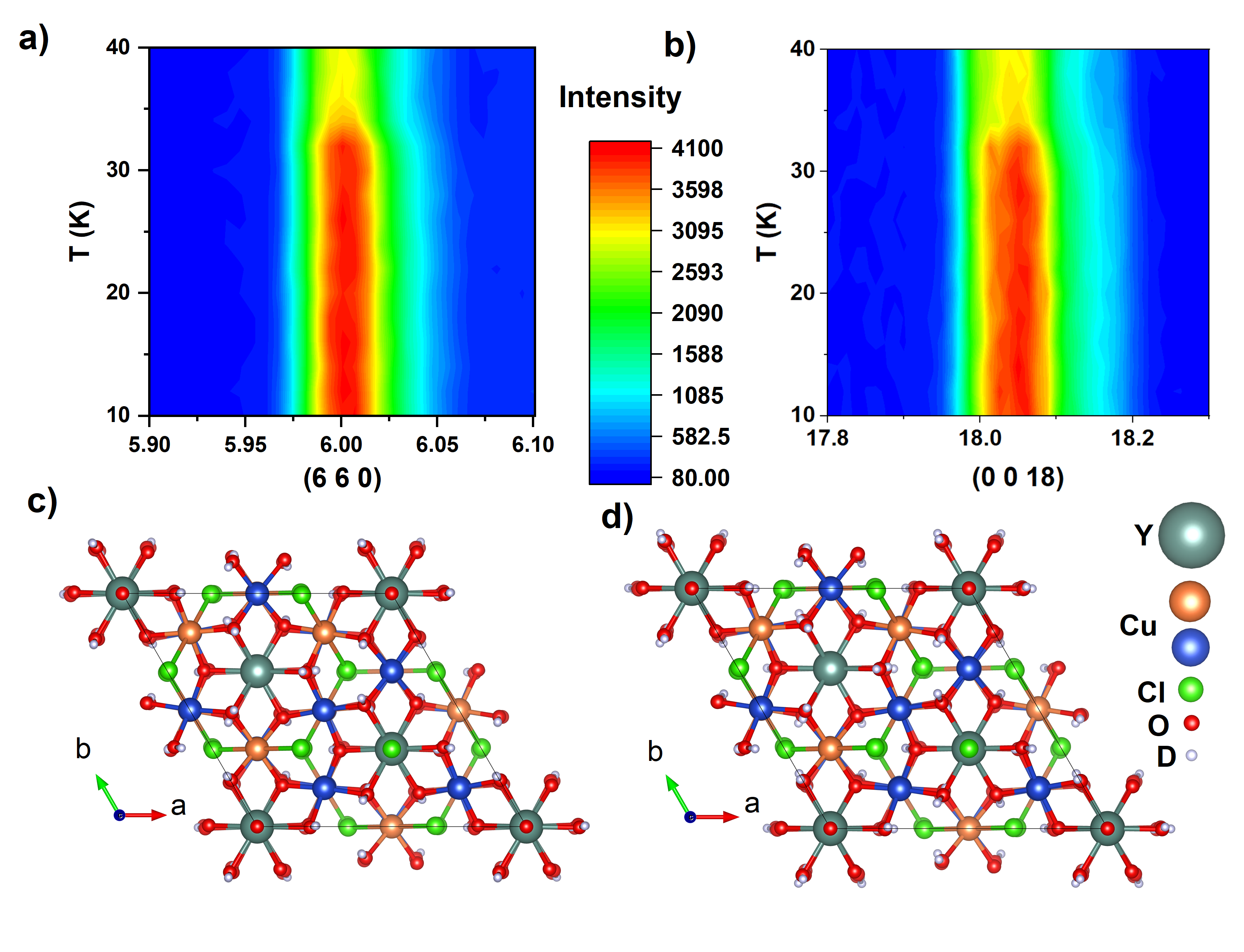}
\par\end{centering}
\caption{Temperature dependence of the neutron diffraction intensity around the nuclear Bragg peaks a) (6 6 0) and b) (0 0 18). Structure viewed along the $c$-axis obtained from neutron single crystal diffraction measurements at c) 40~K and d) 65~mK.}
\label{diffraction}
\end{figure}
The structure of Y-Kapellasite  Y$_3$Cu$_9$(OH)$_{19}$Cl$_8$ with R-3c  was first reported in Ref.~\cite{Puphal2017}, where hydrogen was placed by symmetry arguments. Later, from neutron diffraction on deuterated powder samples, we found that the deuterium atom next to the O1 refines to a distance above 1 \AA, which is so far unaccounted for an O-H bound. Thus in Ref.~\cite{Barthelemy2019}, we dismissed the D1, leading to the revised chemical formula of Y$_3$Cu$_9$(OD)$_{18}$OCl$_8$. To tackle further this issue we then measured the stoichiometry including the hydrogen content with a combination of inductively coupled plasma mass spectroscopy (ICP-OES) and gas extraction, the former with a SPECTRO CIROS CCD and the latter with an Eltra ONH-2000 analyzer. For the determination of oxygen and hydrogen
content, the powder samples were placed in a Ni crucible and clipped and heated, where the carrier gas takes the oxygen and hydrogen out of the sample. The oxygen reacts with carbon, and CO$_2$ is detected in an infrared cell, while hydrogen is detected by a thermal conductivity cell. Each measurement is repeated three
times and compared to a standard. The quoted error bars provide
the statistical error. We thus measured three pieces of a hydrated single crystal and a part of a powder sample of Ref.~\cite{Barthelemy2019}. The results of: Y$_{3.00(3)}$Cu$_{9.06(9)}$O$_{19.0(2)}$H$_{18.7(3)}$Cl$_8$ for the crystals and Y$_{3.00(3)}$Cu$_{9.07(9)}$O$_{19.3(2)}$H$_{18.8(3)}$Cl$_8$ for the powder sample show a high enough hydrogen content to assume the ideal stoichiometry of Y$_3$Cu$_9$(OH)$_{19}$Cl$_8$. Notably we have also seen vibrational O-H modes that exist only for electrical field parallel to the kagome plane solely matching to this hydrogen/deuterium position further confirming the presence of the D1 atom. Therefore, from now on, we use the ideal stoichiometry of Y$_3$Cu$_9$(OH)$_{19}$Cl$_8$. 

We performed neutron diffraction on the  2 axis thermal neutron diffractometer D23 instrument at the Institut Laue-Langevin (ILL) using an inclusion-free, deuterated single crystal weighing around 30~mg. Measurements have been performed in the temperature range of 40~K to 0.065~K using constant incident neutron wavelength $\lambda$=1.27~\AA. We find the R-3c structure that was reported in Ref.~\cite{Puphal2017} to be stable over the whole temperature range. The previously mentioned D1 position can be reasonably refined in all cases to a similar position and even a release of the occupation is possible. We find in all cases an unusually large distance of d$_{O-D}=1.3$~\AA, which was nevertheless considered as mentioned above. The transition around 33~K observed in specific heat and thermal expansion is reflected here in an increase of the Bragg peak intensity by approximately 30\% as shown in Fig.~\ref{diffraction} a) and b). Yet, it produces no change in the crystallographic cell, as the crystallographic axes $a,b$ and $c$ remain unaltered in the investigated temperature range. We find in Rietveld refinements a sudden jump around 33~K for certain atomic positions, which in absolute numbers remain however marginal. The Rietveld refinement results are summarized in table~\ref{Rietveld} and selected atomic positions are plotted normalized to the 40~K values versus temperature in Fig.~\ref{positions} to visualize the sudden changes around 30~K (see also~\footnote{See Supplemental Material at [URL] for the Crystallographic Information Files corresponding to the refined Y-Kapellasite Y$_3$Cu$_9$(OH)$_{19}$Cl$_8$ structure at 40~K, 20~K, 8~K and 0.065~K.}). 
These changes lie below 1\% of the nominal value, except for deuterium and thus are hardly visible in Fig.~\ref{diffraction} c) and d) depicting the crystal structure along the $c$ axis at 40~K and 65~mK, but will be relevant for the discussion of our NMR results.

\begin{figure}[b]
\begin{centering}
\includegraphics[width=1\columnwidth]{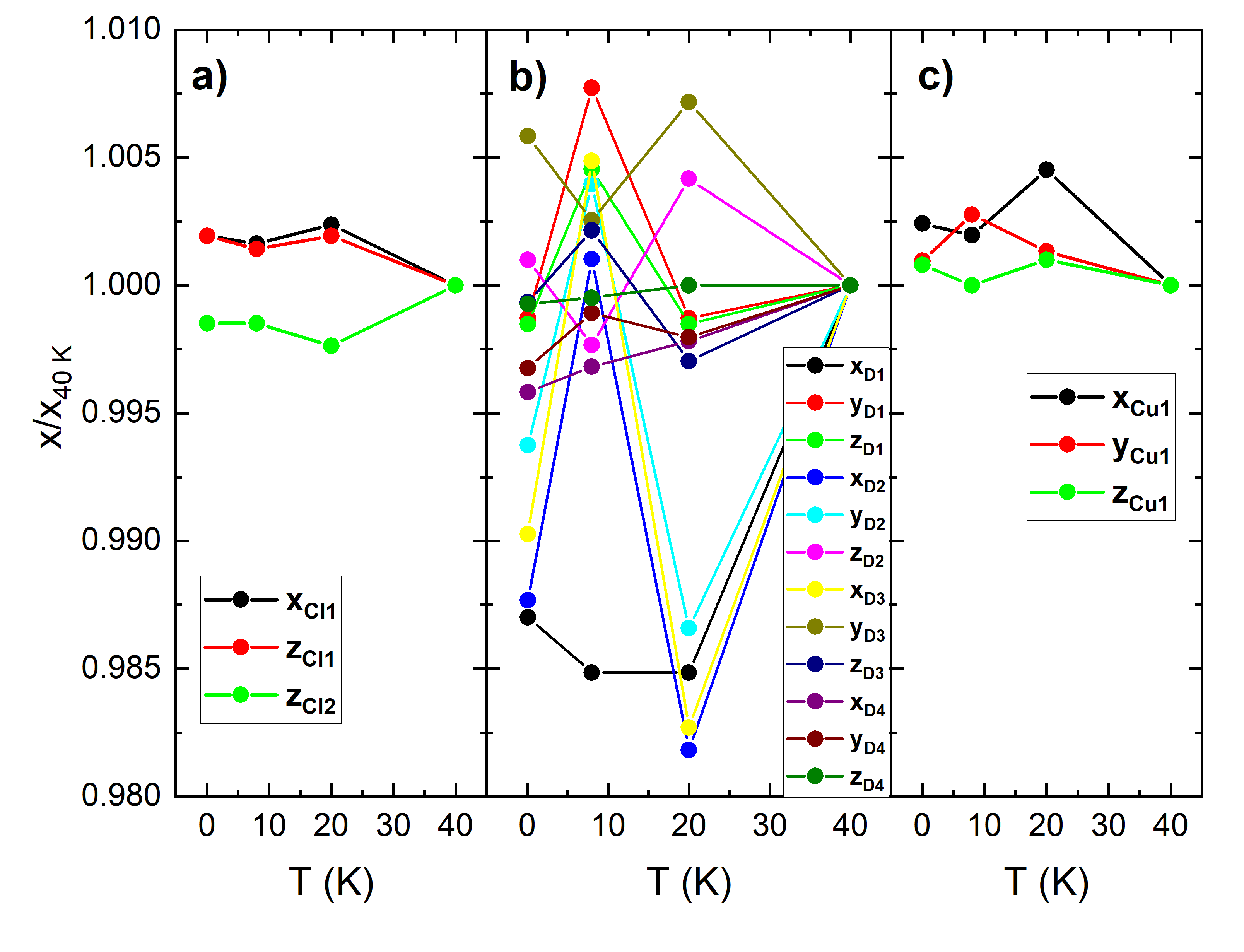}
\par\end{centering}
\caption{Relative atomic position changes versus temperature of a) Chlorine site 1 and 2, b) Deuterium site 1, 2, 3, 4 and c) Cu site 1.}
\label{positions}
\end{figure}
\par At low temperature $T=65$~mK we have searched for magnetic ordering at different (H,K,L) planes but we found no magnetic Bragg peaks probably due to low magnetic moment of Cu$^{2+}$ ions in the ground state. Hence extensive studies of the low temperature regime has been done to investigate the magnetic structure using NMR, $\mu$SR and inelastic neutron scattering, as presented in the following sections.

\section{Nuclear Magnetic Resonance}
We further investigated the structural distortions and magnetic transition using $^{35}$Cl NMR on a $\sim 30$~mg single crystal of Y-Kapellasite Y$_3$Cu$_9$(OH)$_{19}$Cl$_8$ with an external magnetic field B$_{ext}$ applied parallel to the $c$-axis, i.e. perpendicular to the large facet of the sample (see Fig.~\ref{growth}). The spectra are measured from room temperature down to 5~K by sweeping the frequency in a fixed field B$_{ext}=7.553$~T which corresponds to the reference frequency $\nu_0=31.510$~MHz. At lower temperatures, the broader spectra were recorded by sweeping the field with a fixed irradiation frequency $\nu_0$.

\begin{figure}[b]
     \centering
     \includegraphics [width=1\columnwidth] {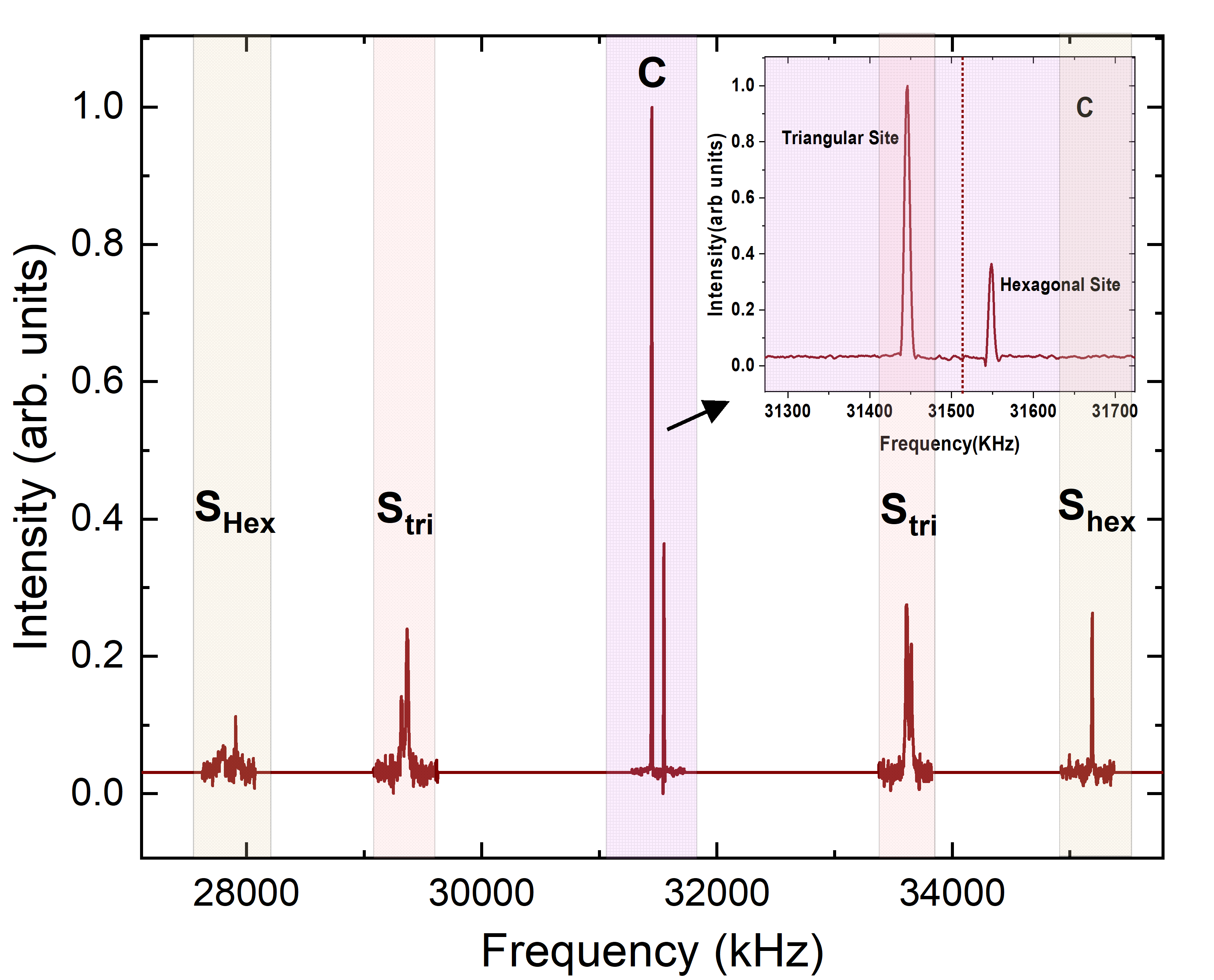}
     \caption{Full NMR spectrum of $^{35}$Cl at 100~K with the configuration $B_{ext}\parallel c$ which shows 2 central lines (C) of Cl1 (triangular site) and Cl2 (hexagonal site) and the 2 corresponding pairs of satellites (S$_{tri}$ and S$_{Hex}$). Inset: enhanced view of the central lines (C). The dashed vertical line indicates the reference frequency $\nu_0=31.510$~MHz}
     \label{NMR-1}
 \end{figure}

The spectrum measured at 100~K, far above the magnetic transition and the structural distortions, is shown in Fig.~\ref{NMR-1}. We observe two distinct central lines (see inset of Fig.~\ref{NMR-1}) corresponding to the two crystallographic Cl sites with intensity ratio close to the expected 3:1.  There are indeed 18 Cl1 (resp. 6 Cl2) sites per unit cell located in between the kagome layers, close to the center of each triangle (resp. hexagon) of the kagome structure. For clarity in this section we denote the Cl1 and Cl2 sites as the triangular ($tri$) and hexagonal ($hex$) sites respectively. 

Since $^{35}$Cl possesses a nuclear spin $I=3/2$ with a finite quadrupolar moment, the three allowed transitions between the Zeeman-split adjacent nuclear levels are modified by the quadrupolar interaction with the surrounding electric charges. We therefore observe in the full spectrum three NMR lines per Cl site, one central line ($-1/2 \leftrightarrow 1/2$ transition) and two satellite ones ($-3/2 \leftrightarrow -1/2$ and $1/2 \leftrightarrow 3/2$ transitions). 

At first order in perturbation,  the frequency difference between two satellite lines of a given $\alpha=tri$ or $\alpha=hex$ Cl site arising from the quadrupolar interaction is given by~\cite{cohen1957quadrupole}
\begin{equation}\label{1st_order}
\Delta \nu_\alpha^{(1)}=\nu^{\alpha}_Q (3\cos^2{\theta_{\alpha}}-1-\eta_{\alpha} \sin^2{\theta_\alpha}\cos 2\phi_\alpha)
 \end{equation}
 where  $\theta_\alpha$ and $\phi_\alpha$ are the polar and azimuthal angles defining B$_{ext}$ in the local frame of the principal axes of the electric field gradient tensor (EFG),  $\nu_Q^\alpha$ is proportional to the quadrupolar moment and the largest eigenvalue of the EFG, therefore reflecting the strength of the quadrupolar interaction, and the asymmetry parameter $\eta_\alpha$ reflects the departure of the EFG from cylindrical symmetry.   

Because of the high symmetry of the hexagonal Cl site, the EFG at this site is axially symmetric ($\eta_{hex}=0$) along the $c$ axis and, under the condition of the experiment $B_{ext} \parallel c$, $\theta_{hex}=0$. For this site, the two satellites  are maximally separated by $2\nu_Q^{hex}$ and we measure directly from Fig.~\ref{NMR-1} $\nu_Q^{hex}=3640(5)$~kHz. The triangular Cl site is less symmetric. We used the structure determined at 40~K in section IV by neutron diffraction to compute the EFG at this site with a point charge approach. From the calculated parameters $\theta_{tri}=8.3^\circ$, $\phi_{tri}=123.0^\circ$, $\eta_{tri}=0.33$ and the measured distance between satellites, we obtain with Eq.~\ref{1st_order} the value $\nu_Q^{tri}=2180(20)$~kHz. Note that if B$_{ext}$ is not strictly applied along the symmetry axis $c$, the triangular sites become non-equivalent because of a different orientation of their EFG with respect to B$_{ext}$. The minute splitting of the triangular site satellite lines in Fig.~\ref{NMR-1} 	reflects such a small misalignment $\pm 1.2^\circ$ of our crystal in the experiment.

The quadrupolar interaction shifts the position of the central line only at second order in perturbation~\cite{baugher69}:
\begin{equation}\label{2nd_order}
\nu^{(2)}_\alpha=-\frac{(\nu^{\alpha}_Q)^2}{2\nu_0} f(\theta_\alpha, \phi_\alpha, \eta_\alpha) 
 \end{equation} 
where, for the simplest case of an axially symmetric EFG, $f$ reads 
\begin{equation}\label{2nd_order_f}
f(\theta, \phi, 0)=\frac{3}{8}(1-\cos^2{\theta})(9\cos^2{\theta}-1). 
\end{equation}
In the present case at 100~K, the second order quadrupolar shift is either $\nu^{(2)}_{hex}=0$ since $\theta_{hex}=0$ or negligible $\nu^{(2)}_{tri}=-4(1)$~kHz. 
The positions of the two central lines
\begin{equation}\label{2nd_order}
\nu^C_\alpha=\nu_0(1+K_\alpha^0+K^s_\alpha)+\nu^{(2)}_\alpha
 \end{equation}
are then rather set by the first term of magnetic origin, where $K_\alpha^0$ is the temperature-independent orbital shift and $K^s_\alpha$ the spin shift arising from the polarization of the unpaired electron of the neighbouring Cu$^{2+}$ ions and therefore proportional to the spin susceptibility. From the plots of $\nu^C_\alpha$ versus the susceptibility $\chi_{c}$ measured along the $c$ direction in the range 100-300~K (not shown), we extract for the configuration B$_{ext} \parallel c$ the hyperfine constants $A_{hex}=0.245(3)$~T/$\mu_B$ and $A_{tri}=-0.426(16)$~T/$\mu_B$  and the orbital terms $K_{hex}^0=120(10)$~ppm and $K_{tri}^0=52(10)$~ppm.  

\begin{figure}[t]
     \centering
     \includegraphics[width=1\columnwidth] {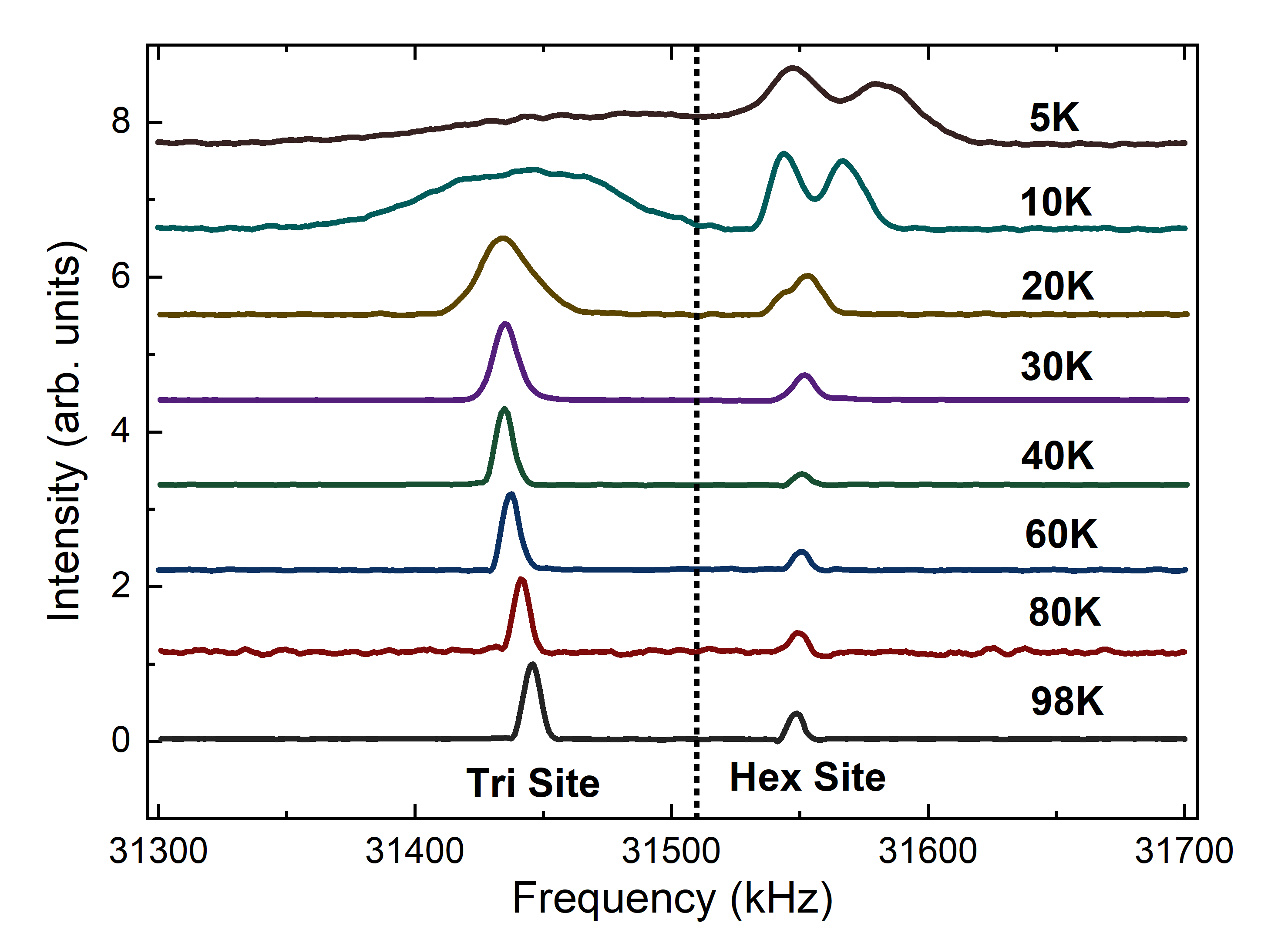}
     \caption{Evolution of the central lines of the $^{35}$Cl NMR spectra with temperature. The spectra are shifted vertically for clarity. Note the change of the width and the splitting of the line below 40~K which become prominent below 20~K. The dashed vertical line indicates the reference frequency.}
     \label{NMR-2}
 \end{figure}

Figure~\ref{NMR-2} shows the evolution of the central lines upon cooling down to 5~K.  Marked changes of the spectral shape are observed below 40~K where the structural distortions take place: the triangular site line broadens significantly and the hexagonal one splits. Interestingly, although the changes appear in the range 30-40~K, they really get prominent only below 20~K where a second peak is observed in thermal expansion. 

From the neutron diffraction study, the structural distortion at 33~K does not affect the crystal symmetry and maintains only two distinct Cl sites. It mainly involves small displacements of the protons, which naturally modify the EFG and therefore the positions of the lines but cannot account for line broadening or line splitting. Therefore, $^{35}$Cl NMR reveals more complex distortions with local symmetry breaking but no spatial correlation so that the average symmetry as detected by neutron diffraction  is preserved.   In particular, the splitting of the hexagonal site line implies the occurrence of at least two locally non-equivalent crystallographic sites and at least one with a finite tilt $\theta_{hex}$ of its EFG so that $\nu^{(2)}_{hex}$ does not vanish in equation~\ref{2nd_order}. From the measured splitting and using Eq.~\ref{2nd_order}, we estimate $\theta_{hex} \sim 5^\circ$ at 20~K and $\theta_{hex} \sim 10^\circ$ at 10~K. 

\begin{figure}[t]
     \centering
     \includegraphics [width=1\columnwidth] {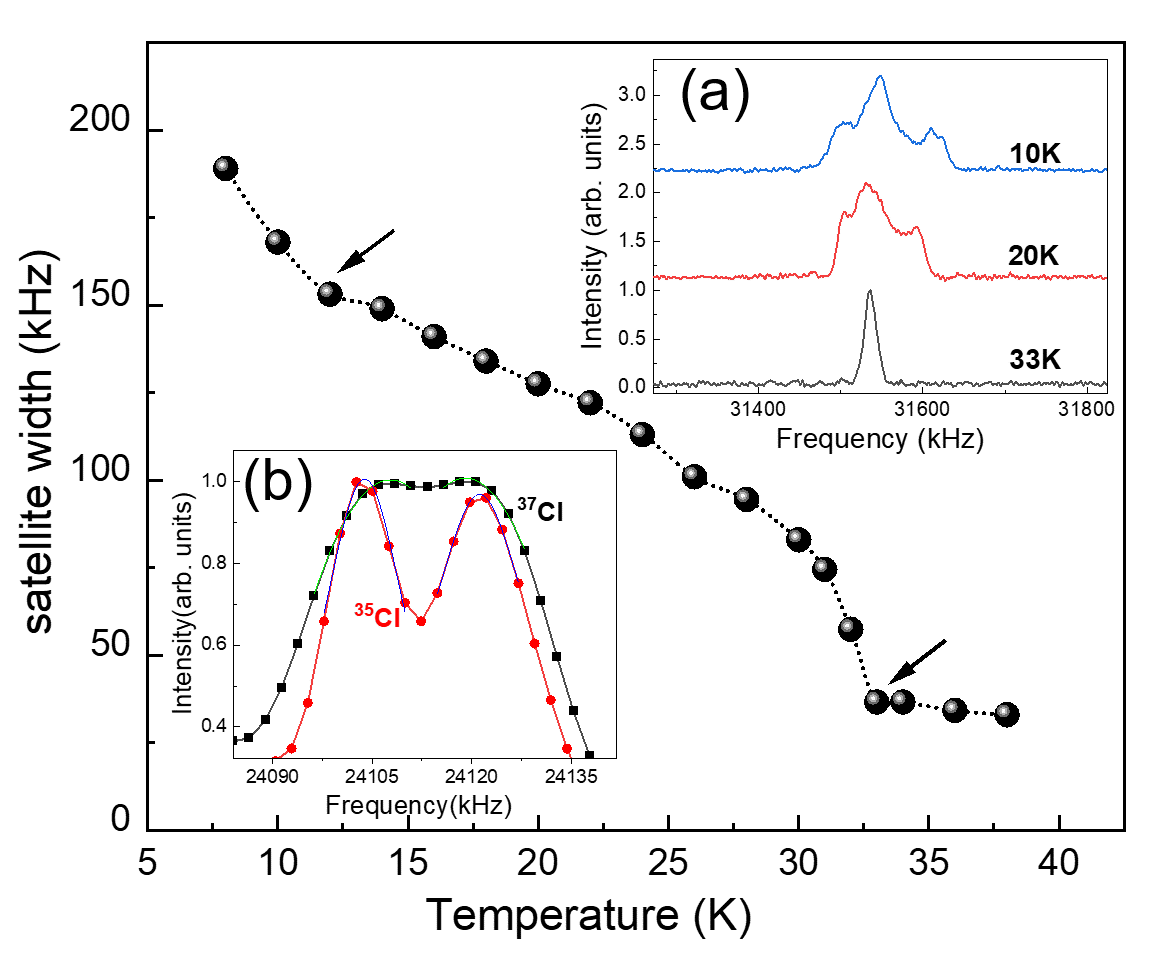}
     \caption{Hexagonal satellite linewidth evolution with temperature which indicates 2 structural changes (black arrows). Inset (a) depicts the line shape changes occurring below 33~K for the hexagonal satellite. Inset (b) shows the hexagonal NMR central lines for $^{37}$Cl and $^{35}$Cl at 10~K.}
     \label{NMR-3}
 \end{figure}

These changes of the local electrostatic environment of Cl affect even more clearly, at first order, the satellite lines. As shown in the inset a) of Fig.~\ref{NMR-3}, the hexagonal satellite acquires a complex structure below 33~K pointing at even more than two nonequivalent sites.  In order to determine the onset of the structural transition, we have measured the width of this satellite versus temperature as plotted in Fig.~\ref{NMR-3}. Beside the abrupt change at the $33$~K structural transition evidenced in heat capacity and neutron diffraction experiments, a marked change of the slope is detected at 13(1)~K suggesting a secondary transition as detected in thermal expansion and also clearly in NMR relaxation measurements as discussed further in this section.

 To confirm the structural nature of the 13~K transition we have compared  the $^{35}$Cl and $^{37}$Cl NMR spectra at 10~K at the same reference frequency of 24.000 MHz. We have fitted the split hexagonal site central lines with gaussian functions and compared the frequency differences $\Delta \nu ^i$ between the two peak maxima for the two isotopes $i$ (inset (b) Fig.~\ref{NMR-3}). The ratio $\Delta \nu ^{35}/\Delta \nu ^{37}=1.5(2)$  compares well to the square of the ratio  of the quadrupolar moments of the two isotopes 1.61 as expected from Eq.~\ref{2nd_order}. At variance, in case of a magnetic origin with static internal fields, we would have observed a scaling with the gyromagnetic ratios of the two isotopes, i.e. a ratio $\sim 1.20$.

\begin{figure}[t]
     \centering
     \includegraphics [width=1\columnwidth] {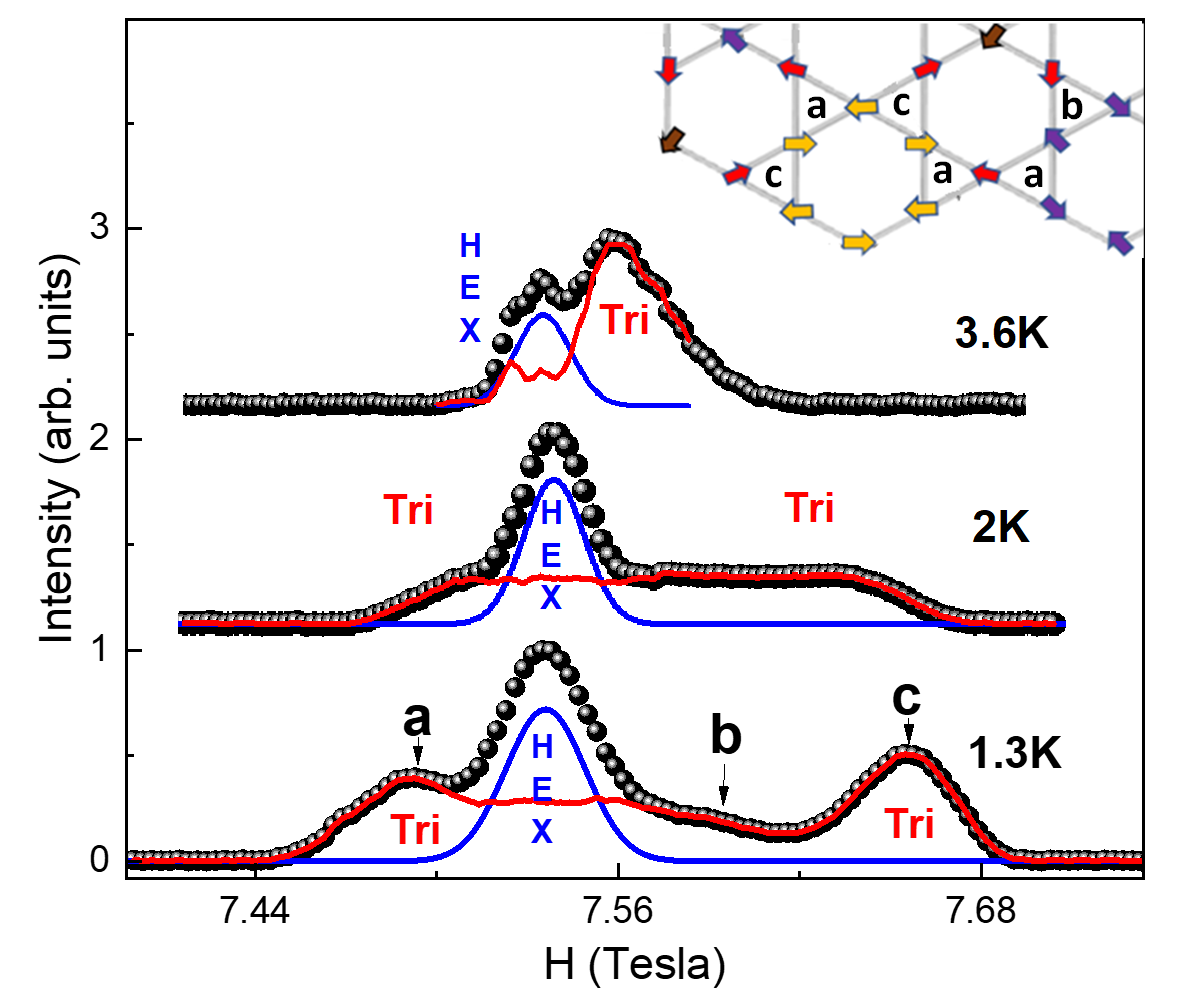}
     \caption{Central line spectra at 1.3~K, 2~K, 3.6~K  and  partial view of the ground state spin texture proposed in Ref.~\cite{hering2022} in the top right corner. The spectra are shifted vertically for clarity. The blue line is a gaussian fit of the hexagonal central line. This contribution is subtracted from the experimental spectrum to expose the triangular site one, shown as a red line.  The triangle configurations \textbf{a}, \textbf{b} and \textbf{c} yield three different dipolar fields at the triangular Cl site. Configurations \textbf{a} and \textbf{c} yield almost equal but opposite dipolar fields whereas configuration \textbf{b} has almost zero dipolar field along the $z$ axis, i.e. the applied field direction.  }
     \label{NMR-8}
 \end{figure}

\begin{figure}[t]
     \centering
     \includegraphics [width=1\columnwidth] {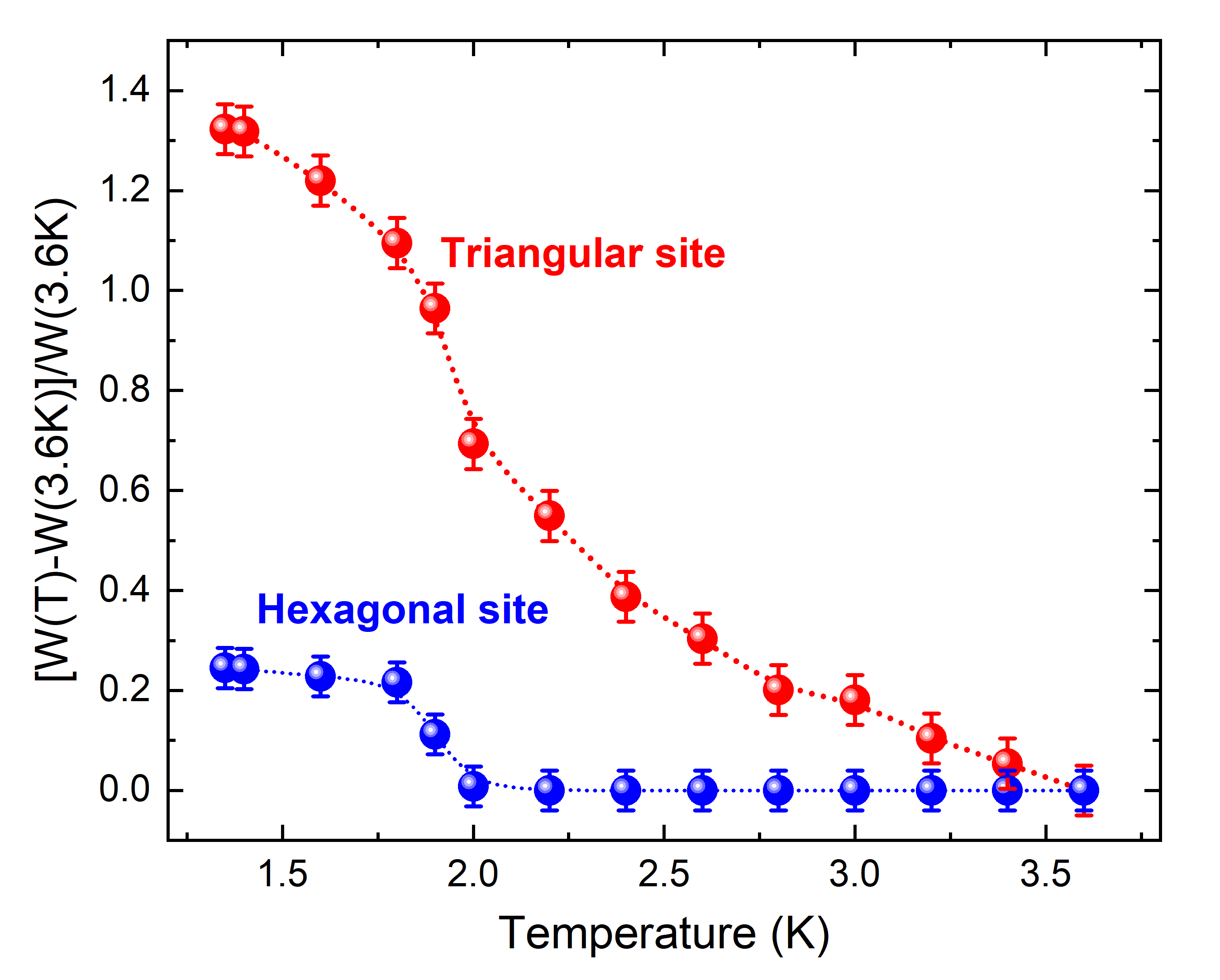}
     \caption{Evolution versus temperature of the linewidth $W(T)$ for the triangular (red dots) and hexagonal (blue dots) sites.}
     \label{NMR-6}
 \end{figure}

At lower temperatures ($T<4.2$~K), the NMR lines broaden significantly, likely as a result of the building up of magnetic correlations, in line with the FC/ZFC divergence observed in this $T$ range in the susceptibility (see Fig.~\ref{growth}). This magnetic broadening blurs the quadrupolar splitting  and down to  2.1~K, we observe only two broad central lines, overlapping partially, for the two different, triangular and hexagonal, chlorine sites  (see Fig.~\ref{NMR-8}). At lower temperatures an abrupt broadening for the triangular site and more moderately for the hexagonal one occurs, which represents the spectral NMR signature of the magnetic transition detected in the bulk, $T_1$ or $\mu$SR measurements. Actually, two triangular site peaks develop on both sides of the hexagonal central line and move away from each other upon cooling down. To quantify these evolutions, we compare in Fig.~\ref{NMR-6} the change of the linewidth $W(T)$  with temperature for the two sites. 

In the recent theoretical study Ref.~\cite{hering2022}, a $(1/3,1/3)$ long range magnetic order has been proposed for this system with a coplanar spin configuration reproduced partially in Fig.~\ref{NMR-8}.
Remarkably, for this predicted spin configuration, the hyperfine field at the hexagonal sites is zero (hexagons with different colors spins) or nearly zero (hexagons with same color spins). At variance, for each triangle, two spins are nearly opposite leaving the third one uncompensated and thus able to yield a finite field at the triangular sites. There are actually three types of triangle configurations labelled \textbf{a}, \textbf{b} and \textbf{c} in Fig.~\ref{NMR-8} corresponding to three different orientations of this uncompensated spin. For an anisotropic hyperfine tensor, we thus expect three different lines for the triangular Cl in the magnetically ordered phase. Although we could not determine experimentally all the components of the hyperfine tensor, we note that assuming a purely dipolar coupling, the shown spin configuration yields one very small and two almost equal but opposite $z$-components of the transferred fields at the triangular Cl site. Thus, the low temperature spectrum supports at least qualitatively the theoretically proposed order, both for the triangular and hexagonal Cl sites. 

\begin{figure}[t]
     \centering
     \includegraphics [width=\columnwidth] {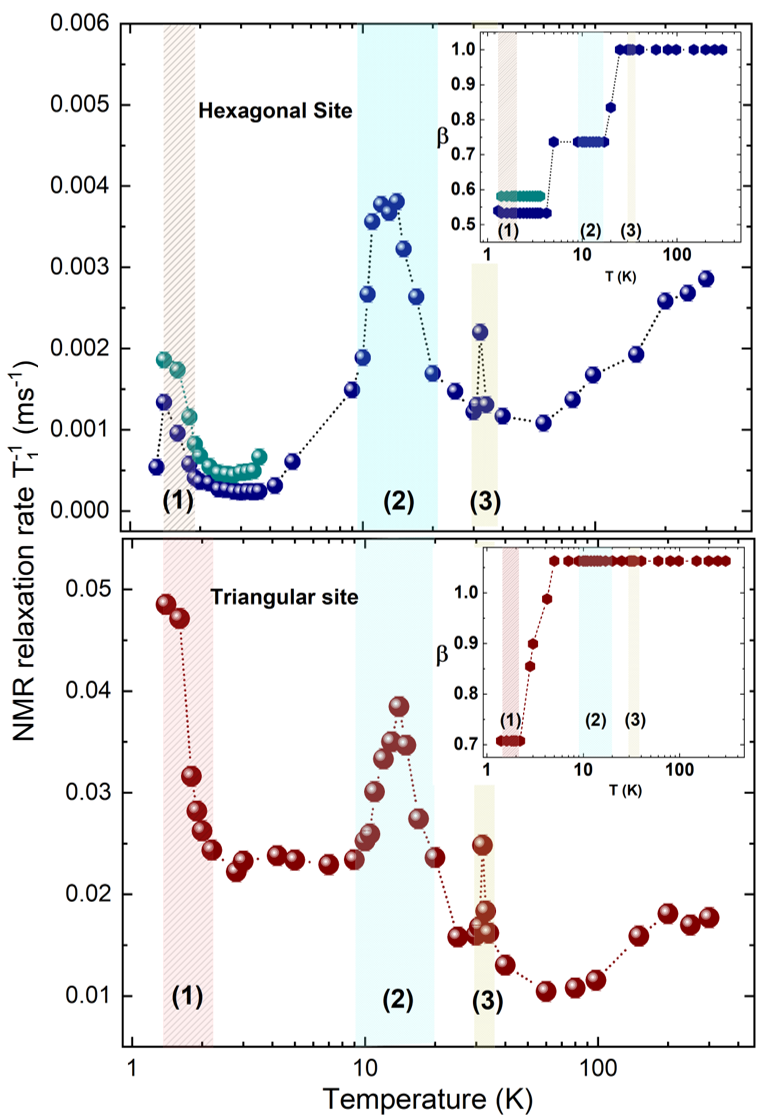}
     \caption{Plots of $^{35}$Cl NMR relaxation rate $T_1^{-1}$ versus temperature measured at the hexagonal (top) and triangular (bottom) central line positions. The two structural transition regimes (2) and (3) and the magnetic one (1) are highlighted. The $T_1$ distribution parameter $\beta$ is plotted in the inset. Below 5~K, when the NMR lines of the two sites overlap, $T_1$ was also measured at the isolated hexagonal satellite position (dark cyan dots) and shows the same behaviour as the central line.}
     \label{NMR-4}
 \end{figure}

 Figure~\ref{NMR-4} shows the evolution with temperature of the spin lattice relaxation rate ($T_1^{-1}$) measured using the saturation recovery sequence at both Cl sites. Being quadrupolar active, Cl nuclei probe both the magnetic and structural fluctuations. In the temperature evolution of $T_1^{-1}$ the structural distortions at $\sim$33~K and $\sim$13~K  discussed earlier are clearly observed as a sharp and broad peak respectively, together with the magnetic transition around $\sim$ 2.1~K.
 
 To extract the $T_1$ values, the recovery curves of the nuclear magnetization $M(t)$ at each temperature were fitted to~\cite{suter1998mixed}
 \begin{equation}\label{eq-1}
M(t)=M_{sat} \left( 1-  a e^{- \left( \frac{t}{T_1} \right)^\beta} - b e^{- \left(\frac{6t}{T_1} \right)^\beta}   \right)
 \end{equation}
where $a+b=1$, $a=0.1$ and $0.23$ respectively for the hexagonal and triangular sites, and $\beta \leq 1$ is a phenomenological stretching parameter used to account for a possible distribution of $T_1$~\cite{Choi21}.  The evolution of $\beta(T)$ for both sites is shown as insets in Fig.~\ref{NMR-4}. While a single value of $T_1$ is measured at high temperature ($\beta=1$), an increasingly broader distribution ($\beta<1$) is needed upon cooling down towards the structural and magnetic transitions, likely pointing at the building up of spatial correlations. During the transition regimes, we could fix the value of $\beta$ with no noticeable reduction of the fit quality, to avoid artefacts in the determination of the peaks of the correlated $T_1^{-1}$ quantity.

\section{Muon spin relaxation}

To elucidate the nature of the ground state of Y- Kapellasite, we performed a \(\mu\)SR experiment on the GPS and DOLLY instruments at the PSI muon facility from the paramagnetic regime (\(>\)20~K) down to 0.28~K. The 100\% spin polarized muons are implanted at time zero in the sample, where they stop at the most electronegative positions. The muon spins evolve in the local fields until the muons decay into positrons, emitted preferentially in the last direction of the muon spins after a mean life time of 2.2 \(\mu\)s. By collecting the positrons emitted forward and backward with respect to the muon beam direction, the muon decay asymmetry, proportional to muon spin polarization can be recorded versus time and allows to characterize the internal fields and their fluctuations~\cite{1,2,hillier2022muon}.

\begin{figure}[t]
     \centering
     \includegraphics [width=1\columnwidth] {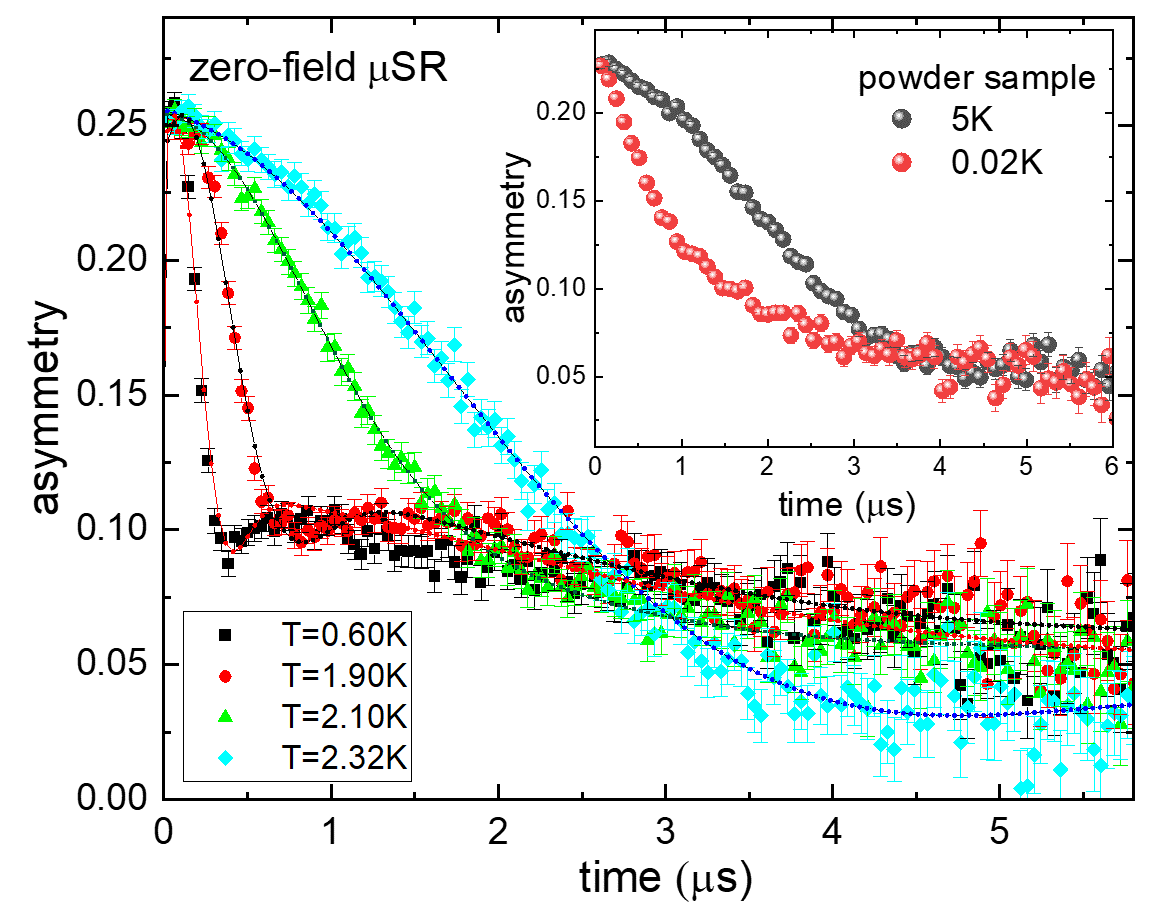}
     \caption{Time evolution of the zero-field (ZF) muon decay asymmetry at some selected temperatures in single crystal samples. Lines are fit to Eq.~\ref{eq-5}. Inset: powder sample data from Ref.~\cite{Barthelemy2019}.}
     \label{fig:1}
 \end{figure}

For this $\mu$SR experiment, we used two coaligned phase pure single crystals of sizes $\sim 3\times3\times1$~mm$^3$ with their $c$ axis parallel to the initial muon spin direction. The muon decay asymmetry in zero external field (ZF) is depicted at some selected temperatures in Fig.~\ref{fig:1}. At low temperature, the asymmetry evidently differs from previous measurements~\cite{Barthelemy2019} on a powder sample shown in the inset, with a much faster relaxation and a characteristic damped oscillation signaling static internal fields.
 
 Above 2.5~K, the relaxation hardly depends on temperature nor on the nature of the sample, powder or single crystal. The electronic spins are  fast fluctuating in their paramagnetic regime, and the relaxation is dominated by the quasi static, weak and random, nuclear fields. Following the model used for the powder sample~\cite{Barthelemy2019}, we fitted the muon decay asymmetry at high temperature to 
 \begin{equation}\label{eq-2}
 a_0 P_{para}(t)=a_0 \left[ f P_{OH}(t) + (1-f)KT_{\Delta_{Cl}}(t) \right]\\
 \end{equation}
 where
\begin{multline}\label{eq-3}
P_{OH}(t) = e^{-\frac{ \Delta^2_{OH}t^2} { 2}} {\bigg[}\frac{1}{6}+\frac{1}{3}\cos{{\bigg(}\frac{\omega_{OH}t}{2}{\bigg)}}+\\
\frac{1}{3}\cos{{\bigg(}\frac{3\hspace{1mm}\omega_{OH}t}{2}{\bigg)}}+\frac{1}{6}\cos{{\bigg(}{\omega_{OH}t}{\bigg)}}{\bigg]}.\
\end{multline}
The latter $P_{OH}(t)$ stands for the formation of \(\mu\)OH complexes with a pulsation $\omega_{OH} = \frac{\hbar \mu_0 \gamma_\mu\gamma_H}{4\pi d^3}$ which depends on the $\mu$-H distance $d$, the gyromagnetic ratios $\gamma_H=267.513 $ Mrad/s/T and $\gamma_\mu=851.616 $ Mrad/s/T respectively for protons and muons, and a gaussian broadening $\Delta_{OH}$ due to the other surrounding nuclear spins. The second term in Eq.~\ref{eq-2} is a static Kubo-Toyabe relaxation standing for the minority fraction $1-f$ of muon stopping sites, likely close to the Cl$^{\textbf{-}}$ ions, where they experience a gaussian distribution of static nuclear fields with a width $\Delta_{Cl}$. The parameter $a_0=0.255$ is the initial muon decay asymmetry in our experimental conditions. 

The nuclear relaxation parameters in Eq.~\ref{eq-2} and Eq.~\ref{eq-3} evaluated by fitting the 
ZF muon polarisation at 20~K are presented in table~\ref{Tab:1}.  \\
\begin{table}[h]
\centering
     \begin{tabular}{|c|c|}
\hline
  Static parameters        & Y$_3$Cu$_9$(OH)$_{19}$Cl$_8$\\
\hline
    $f\%$   &  75.00$\pm$1.00\\
\hline
 $\omega_{OH}$~(Mrad.s$^{-1}$) &0.55$\pm$0.02\\
\hline
         d~(\AA) & 1.63$\pm$0.02\\
\hline
         $\Delta_{Cl}~(\mu s^{-1})$ & 0.09$\pm$0.02\\
\hline
      $\Delta_{OH}$~($\mu$s$^{-1}$) & 0.208$\pm$0.02\\
\hline
      \end{tabular}
\caption{Static nuclear parameters derived from high temperature fit of the ZF asymmetry with Eq.~\ref{eq-2}.}
     \label{Tab:1}
   \end{table} 
   
Upon cooling, the relaxation from the  electronic spins increases progressively and below 2.1~K, at variance with the powder sample~\cite{Barthelemy2019}, a strongly damped oscillation develops, conclusively indicating a magnetic transition in the single crystal samples (Fig~\ref{fig:1}). Below 1.5~K, the asymmetry could be fitted accordingly to 
\begin{equation} \label{eq-4}
    a_0 P_f(t)=a_0 \left[ \frac{2}{3} \cos(\omega_f t+\phi) e^{ -\frac{\sigma^2 t^2}{2}} +\frac{e^{-\lambda_ft}}{3}  \right]
\end{equation}   
which accounts for a magnetically frozen single phase with an average internal field at the muon sites $B_{int}=\omega_f/\gamma_\mu$.  The damping parameter $\sigma$ encodes the width of the distribution of these internal fields. The last exponential term accounts for residual fluctuations in the frozen phase. At base temperature $T=0.28$~K, we obtain $B_{int}=8.6$~mT which is almost twice lower than the lowest internal fields previously reported in the closely related $x=0$ variant of YCu$_3$(OH)$_6$O$_x$Cl$_{3-x}$~\cite{Barthelemy2019,zorko2019coexistence}. Although there is an uncertainty about the exact position of the muons, from this dipolar field value, we calculate a strongly reduced ordered magnetic moment for the Cu$^{2+}$ $\sim 0.075$ $\mu_B$ using  the majority \(\mu\)OH complex forming sites. Besides, the use in Eq.~\ref{eq-4} of the coefficient $2/3$ and $1/3$ for the oscillatory and non-oscillatory components in our single crystal, expected for a spin-glass like state, suggests a rather disordered ground state. The strong damping  $\sigma / \gamma_\mu=3.2$~mT compared to $B_{int}$ also points to a rather short range order and may simply reflect the absence of magnetic correlations in the $c$ direction, as shown by inelastic neutron scattering (see section VII).

\begin{figure}[t]
     \centering
     \includegraphics [width=1\columnwidth] {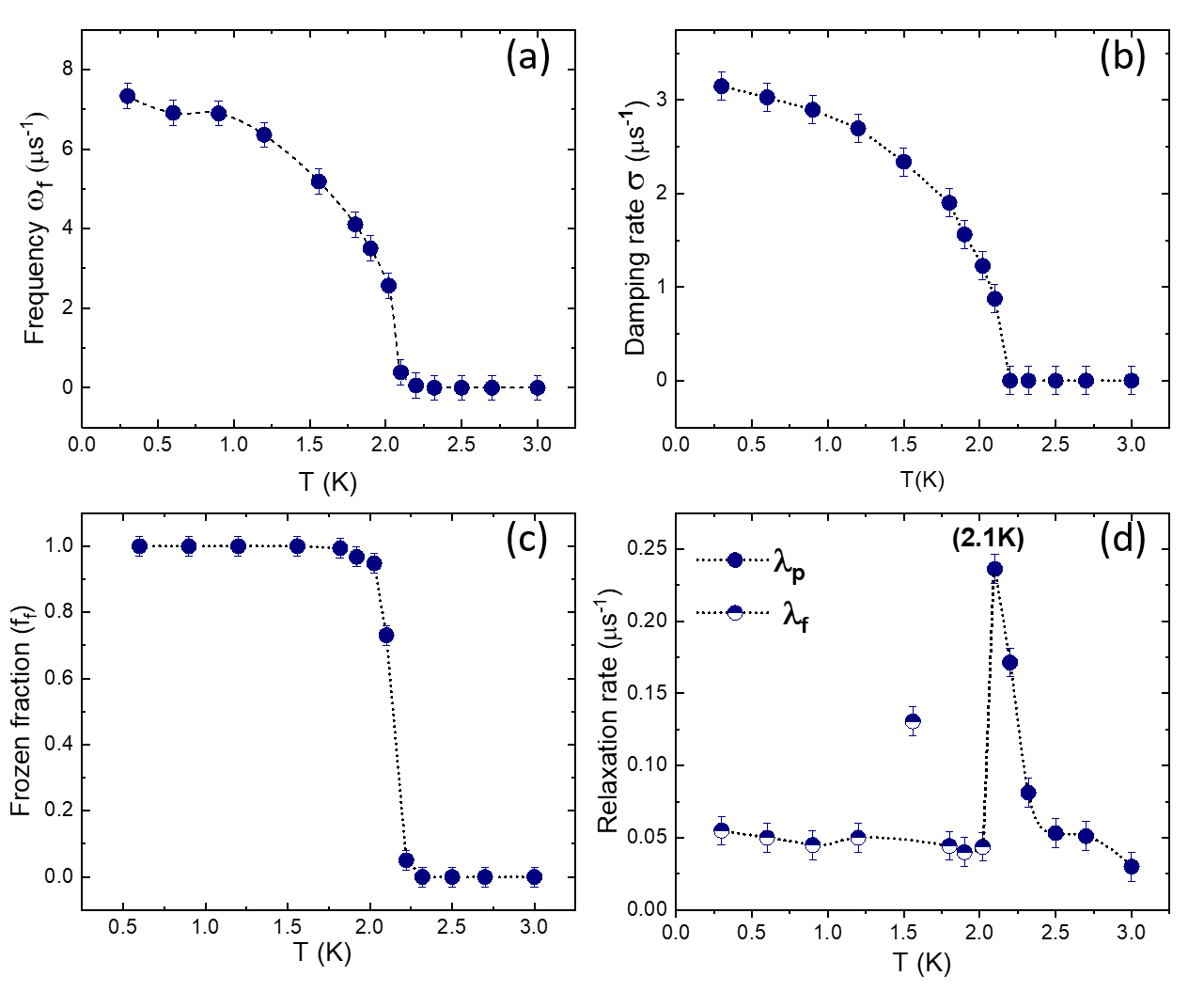}
    \caption{Temperature evolution of the parameters used in Eq.~\ref{eq-5} to fit the ZF $\mu$SR data: a) frequency $\omega_f$ reflecting the internal field magnitude, b) damping rate $\sigma$, c) fraction $f_f$ of the frozen phase and d) relaxation rates in the paramagnetic ($\lambda_p$) and frozen ($\lambda_f$) phases.}
     \label{fig:2}
 \end{figure}

In order to fit the asymmetry on the whole temperature range, we combined both equations \ref{eq-2} and \ref{eq-4} in
\begin{equation} \label{eq-5}
a_0 P(t)=a_0 \left[ f_f P_f(t) + (1-f_f) P_{para}(t) e^{-\lambda_pt} \right] 
\end{equation}
with a switching parameter $f_f$ that tracks the frozen volume fraction from fully ordered for $f_f=1$ to fully paramagnetic for $f_f=0$. All the nuclear parameters are kept constant, so that the only varying parameters shown in Fig.~\ref{fig:2} are the frozen fraction, the oscillation frequency and its damping rate, and the relaxation rate $\lambda_p$ or $\lambda_f$ for the paramagnetic or the frozen phase. The latter show a clear peak at 2.1~K as expected for the critical slowing down of the fluctuations at the magnetic transition. Surprisingly, there is  a singular point at 1.56K in the relaxation rate plot but no significant changes in other variable parameters at the same temperature. Although this feature demands further confirmation, it is reminiscent of the spin dynamics re-entrance reported in clinoatacamite~\cite{Zheng05} and echoes an anomaly reported in the low temperature magnetization measurements~\cite{Biesner22}.

 \par To conclude, from our $\mu$SR measurements, Y-Kapellasite shows a completely frozen ground state in the case of single crystals in contrast to the fluctuating ground state observed in powder sample. The relaxation rate variation  with temperature indicates a phase transition at $\sim$2.1K  where the internal static fields set in. At 0.28~K, deep in the magnetic phase state, we estimate a strongly reduced frozen moment for the Cu$^{2+}$ of about 0.075~$\mu_B$.

\section{Time-of-flight neutron scattering}
\begin{figure*}
\begin{centering}
\includegraphics[scale=0.25]{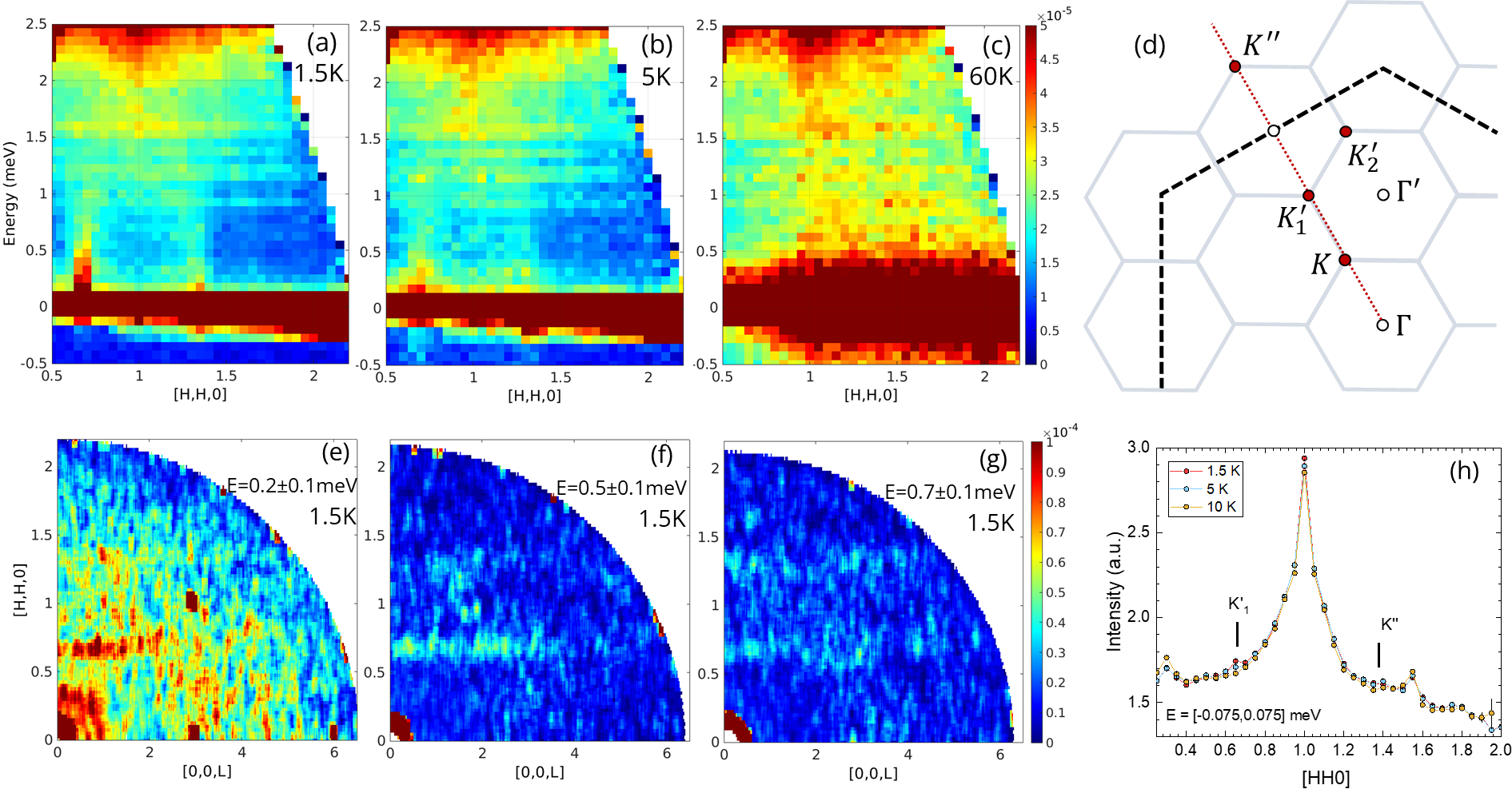}
\par\end{centering}
\caption{(a-c) Evolution in temperature of the dynamical structure factor $S(q,E)$ along the $[H,H,0]$ direction (integrated over $L=[-6,6]$) at $T = 1.55$~K, $T = 5$~K and $T = 60$~K . (d) Brillouin zones boundaries (gray lines) with $K$ and $\Gamma$ points and extended Brillouin zone (black dashed line) in the reciprocal space. The red dotted line represents the $[H,H,0]$ direction. (e-g) Evolution in energy of  $S(q,E)$ in the $(H,H,L)$ plane at $T = 1.55$~K. (h) INS measurement integrated in energy over the elastic resolution showing a weak increase below 5~K at the $K'_1$ and $K''$ positions, characteristic of the $(1/3,1/3)$ magnetic order. }
\label{INS-1}
\end{figure*}
With large deuterated single crystals of Y$_3$Cu$_9$(OD)$_{19}$Cl$_8$ at hand, we also performed inelastic neutron scattering on this anisotropic kagome antiferromagnet in order to probe the magnetic excitations in reciprocal space and to identify the magnetic order~\cite{IN5data}. Inelastic neutron scattering measurements were performed on the IN5 disk chopper time-of-flight spectrometer~\cite{IN5paper} at the Institut Laue-Langevin on a collection of single crystals with a total mass of 0.3g. These crystals were co-aligned in the  $(H,H,L)$ horizontal plane and glued on a thin aluminium plate with the hydrogen-free CYTOP solution (CTL-809M). Measurements runs were performed at a constant temperature with an incident neutron wavelength $\lambda = 4.8$~{\AA} (incident energy E$_i =3.55$~meV, giving a FWHM energy resolution of 0.08~meV) or $\lambda = 2.6$~{\AA} (E$_i =12.1$~meV), while the sample stage was rotated around the direction perpendicular to the beam, to map out the reciprocal space. The detectors efficiency was corrected by a preliminary measurement on a vanadium can.

Magnetic scattering could be identified only below 3~meV, an unexpected feature given the large antiferromagnetic Curie-Weiss temperature of  $\theta_{\rm cw}=-100$~K \cite{Puphal2017}. We thus focused on measurements using our low incident energy of 3.55~meV and followed the temperature evolution of the dynamical neutron structure factor $S(q,E)$. Figure~\ref{INS-1} shows the intensity contour color plots of $S(q,E)$ along the $[H,H,0]$ direction. At 60~K there is a large phonon contribution that substantially reduces at 20~K and below, where nearly vertical columns of intensity become apparent close to $K'_1 = (2/3,2/3,0)$ and  $K''= (4/3,4/3,0)$, with a region of broad intensity located around 2~meV and close to $[1,1,0]$. At our base temperature of 1.5~K, sharp dispersive features emerge from these two $Q$ positions which we identify as spin-wave excitations. They show a pronounced two-dimensional character, as revealed by the presence of rods of scattering intensity along the $[0,0,L]$ direction (Fig.~\ref{INS-1} e-g), and thus justify the 2D Heisenberg kagome antiferromagnet model applied below. From this time-of-flight measurement, one can isolate the elastic contribution by integrating over twice the energy resolution ($\sim 0.15$~meV) only. Figure~\ref{INS-1} h) shows such contribution with the apparition of tiny $q$-resolution limited magnetic Bragg peaks below 5~K located at both $K'_1$ and $K''$ positions. This elastic scattering points to a small ordered fraction of the Cu$^{2+}$ moment, that could be easily missed in a conventional energy-integrated neutron diffraction experiment, and strongly supports a magnetic non-collinear coplanar phase with a $Q = (1/3,1/3)$ ordering wave vector. 

\begin{figure}
\begin{centering}
\includegraphics[width=1\columnwidth]{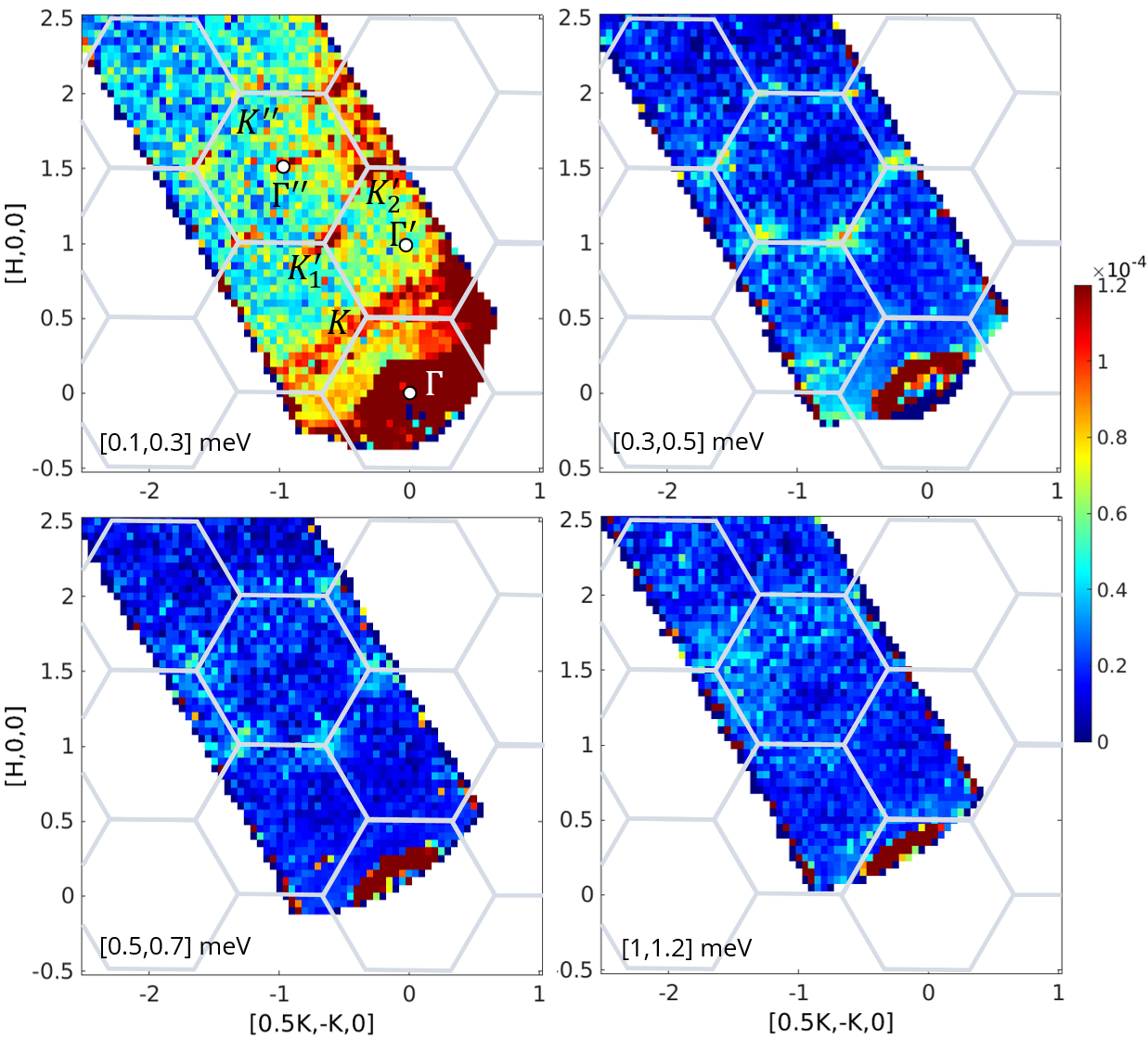}
\par\end{centering}
\caption{Dynamical structure factor $S(Q,E)$ for the $(H,K,0)$ plane measured at $T = 1.55$~K. The four color plot maps show its evolution in energy with the corresponding integrated energy range. Gray lines indicate the Brillouin zones boundaries. Bright spots of intensity are observed at the $K$ points, and corresponds to the dispersion of spin-waves.}
\label{INS-2}
\end{figure}

The inelastic scattering in the $(H,K,0)$ plane is displayed in Fig.~\ref{INS-2}. The energy-integrated cuts show clear intensities located at the $K$ positions of the Brillouin zone, mirroring the sharp spin wave dispersions already observed below 1~meV in Fig.~\ref{INS-1} a). Above 1~meV, the inelastic scattering broadens significantly and extends toward the Brillouin zone center. 

\begin{figure*}
\begin{centering}
\includegraphics[scale=0.59]{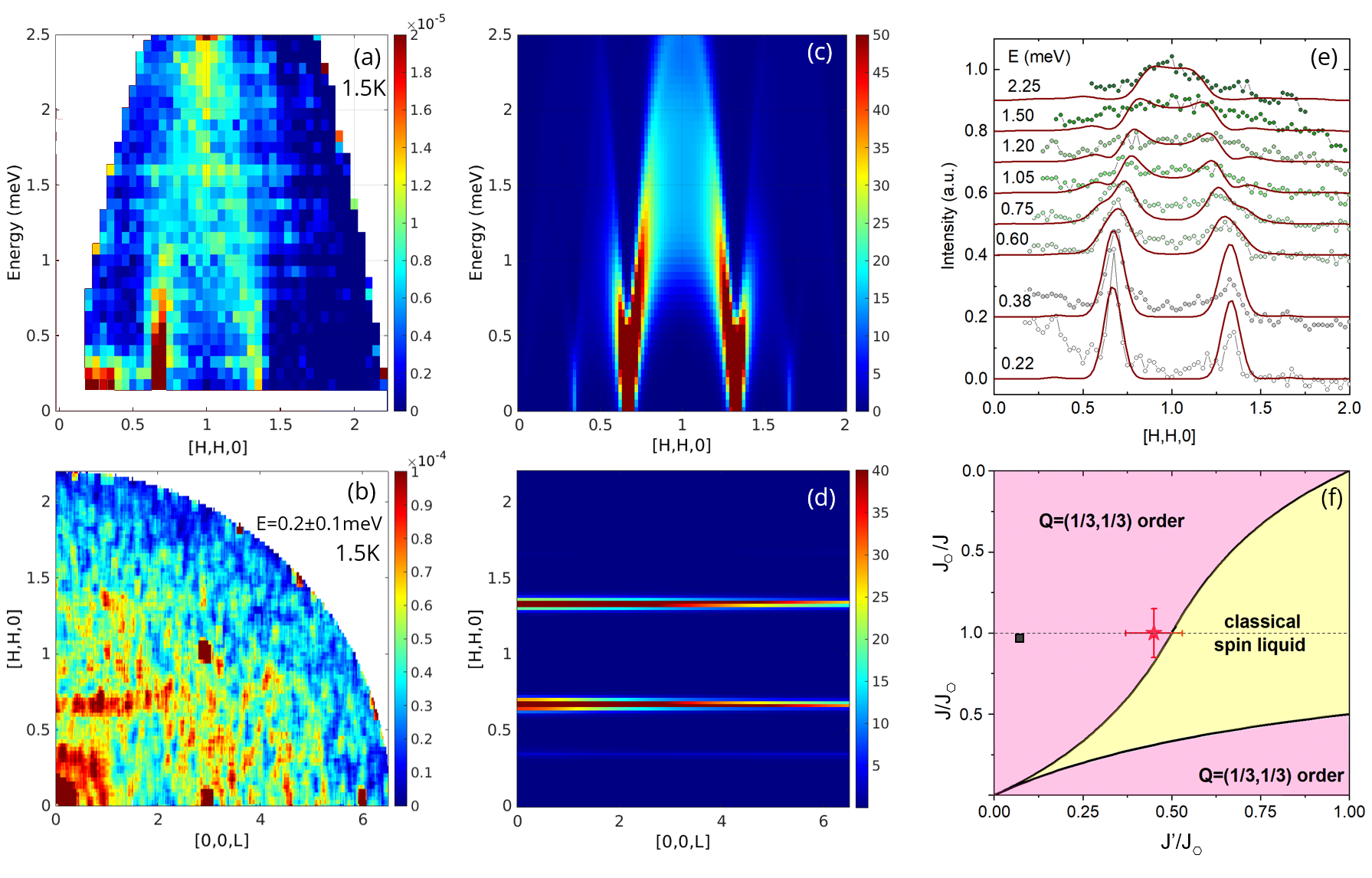}
\par\end{centering}
\caption{Comparison between experimental data (a,b) and spin-waves simulations (c,d) at $T=1.5$~K.  The INS data in (a) measured at 1.5~K was subtracted by a 5~K dataset (see text) to highlight the magnetic contribution. (e) Energy-integrated cuts along $[H,H,0]$ of data shown in (a) and (c) for different energies $E$ with an integration range of 0.15 meV. Symbols are for data and lines for the simulation. They are both shifted vertically  for clarity. (f) Classical phase diagram reproduced from \cite{hering2022}. The black square denotes the position found from DFT calculation \cite{hering2022} while the red star locates Y-kapellasite using the exchange couplings found in our work.}
\label{INS-3}
\end{figure*}

To better isolate the inelastic magnetic contribution present at 1.5~K, we subtracted a background constructed by the replication of the 5~K dataset, integrated in the range $H=[1.6,2]$, along $[H,H,0]$, where the phonon and magnetic contributions appear to be minimal. The result is shown in Fig.~\ref{INS-3} a). The dispersive excitations rising from $K'_1$ and $K''$ are clearly singled out, and then merge into a broad arch extending from 1 to 2.5 meV. Guided by the observation of an in-plane distortion of the kagome lattice, and by the recent ab initio DFT results~\cite{hering2022}, we performed linear spin-wave calculations using the SpinW package \cite{Spinwpackage} of the Heisenberg model with only nearest-neighbors interactions in order to model the excitations on such an anisotropic kagome lattice:
\begin{equation}
\label{Heisenberg}
\mathcal{H} = \sum_{\left\langle i,j \right\rangle} J_{ij} \bm{S_i} \cdot \bm{S_j}\\ 
\end{equation}
\noindent where $J_{ij}$ can take three different values $J_{\hexagon}$, $J$  and $J'$ depending respectively on the bonds $d_1, d_2$ and $d_3$ as defined in Fig.~\ref{growth}.  From DFT calculation, Hering \textit{et al}.~\cite{hering2022} confirmed the relevance of such spin models in the context of Y-Kapellasite using three  antiferromagnetic couplings $J \simeq J_{\hexagon} \simeq 140$~K and $J' \simeq 10$~K. The spin-wave simulations assumed an enlarged unit cell of nine sites, because of the two different copper sublattices, and a coplanar order with a $Q = (1/3,1/3)$ propagation vector, as described in details in Ref.~\cite{hering2022}. The $J =J_{\hexagon} = J'$ case leads to the well-known $\sqrt{3} \times \sqrt{3}$ phase of the classical kagome antiferromagnetic model which fails to reproduce our data when $J>15$~K, due to the maximum of the band observed around 2.5~meV. Given that $\theta_{\rm cw}=-100$~K imposes a higher exchange value, we discard the isotropic case and instead study the excitations for the parameters $J = J_{\hexagon}$ and $J' \neq J$. From the evolution of the spin-wave spectrum with $J$ or $J'$, we find our best model using  $J \simeq J_{\hexagon} = 140$~K and $J' = 63$~K. However, we also find that a single set of interactions can not account for the spread of intensity observed at $[1,1,0]$, from 1 to 2.5~meV. Instead, we consider a $\sim 10$~\% distribution of  $J'$, tentatively attributed to some local disorder, and that could indeed satisfactorily fit our inelastic data, as shown in Fig.~\ref{INS-3} c), d) and e), leading to $J \simeq J_{\hexagon} = 140(10)$~K and $J' = [56,70]$~K. Finally, we note that the intensity observed at low $Q$, near $Q=(1/3,1/3,0)$, and below 0.5~meV (see  Fig.~\ref{INS-3} a) is much broader than predicted by the simulation, and may hint at a more complex ground state in Y-kapellasite with the presence of short-range order and fluctuations. We note that this broad intensity extends mostly around the zone boundaries of the first Brillouin zone (Fig.~\ref{INS-2}) and not close to the boundaries of the \emph{extended} Brillouin zone, as predicted for the classical spin liquid \cite{hering2022}.  

In summary, INS gives evidence for elastic scattering below 5~K, in agreement with the static magnetism revealed by $\mu$SR and line broadening in NMR, consistent with a $Q=(1/3,1/3)$ magnetic ground state. Furthermore, we found a set of magnetic interactions in Y-kapellasite which is in very good agreement with the predictions from DFT and variational Monte-Carlo studies~\cite{hering2022}. The main difference is our experimental value of $J' \simeq 63(7)$~K as compared to the theoretical estimate $J'=10.3(7)$~K. This locates Y-kapellasite in the $Q=(1/3,1/3)$ magnetic order of the phase diagram, but very close to the classical spin liquid phase reported in Ref.~\cite{hering2022}. 

\section{Discussion and Conclusion}

Our experimental investigation of Y$_ 3$Cu$_9$(OH)$_{19}$Cl$_8$ reveals an unusual bidimensional coplanar $Q=(1/3,1/3)$ long range ordered ground state, overall in very good agreement with the theoretical study in Ref.~\cite{hering2022}. In particular, the proposed spin arrangement in the ground state is found to be compatible with the Cl NMR lineshape at low temperature. However, at variance with first principle calculations where the third interaction of the model $J'$ was found to be negligible, we determine experimentally a finite  $J'\lesssim J\hexagon/2$. This finding of a sizeable antiferromagnetic $J'$ seems nonetheless consistent with the Cu-OH-Cu bond angles which vary only by 4.6$^\circ$ among the three different bonds. The smallest angle $113.1^\circ$ corresponding to $J'$ remains significantly larger than 105$^\circ$ which is known to yield small $\sim -15$~K ferromagnetic interaction in Zn-kapellasite~\cite{Fak12}. 

Besides, to reproduce the broad magnetic excitations in INS, a distribution $\sim 10\%$ of the $J'$ interaction was imposed. This suggests some kind of disorder. Due to the large difference in the ionic radii of Y$^{3+}$ and Cu$^{2+}$ we expect no significant intersite mixing at variance with Zn-kapellasite or other Zn and Cu based kagome materials and indeed we found no evidence of such disorder in the magnetic lattice above 33~K, neither in neutron diffraction nor in NMR which on the contrary exposes well resolved narrow lines. Therefore, the distribution of $J'$ may arise only at low temperature, as a result of the structural instabilities observed at 33~K and 13~K. The slightly different local environment detected below 33~K by Cl NMR would then reflect in the moderate distribution of the interaction, while preserving on average the high temperature magnetic model, since the symmetry and lattice parameters remain unchanged on the whole temperature range, down to 65~mK. 

The experimentally determined interactions locate Y-kapellasite in the classical phase diagram close to the boundary between the (1/3,1/3) long range  ordered and the "classical spin liquid" phases (see Fig.~\ref{INS-3} f). Although the effect of quantum fluctuations has not been investigated so close to the phase boundary, we may anticipate that they are responsible for the strongly reduced value of the Cu$^{2+}$ moments $\lesssim 0.1\mu_B$, which could not be detected in standard neutron diffraction but yield static internal fields detected both in NMR and $\mu$SR. The proximity to a phase boundary may also help understanding the striking difference in the ground states of the present large  single crystal samples and formerly studied polycrystalline ones~\cite{Barthelemy2019}. Indeed in the latter, the moments were found to remain fluctuating in the ground state and also no structural transition was detected. We may assume that some additional disorder in the polycrystalline samples or slightly different interactions, due to the absence of structural transition, destabilize the fragile ordered state observed in large single crystals. We note that such a sensitivity to the sample crystallinity seems to be a common feature of anisotropic kagome compounds like volborthite~\cite{bert05,Ishikawa2015} or vesigneite~\cite{Quilliam11,Yoshida13,Boldrin2016}. The fragility of the ground state in all these compounds seems to originate from a subtle interplay between structure and magnetic frustration.


Beyond the proposed anisotropic nearest neighbor model, and despite its apparent success to reproduce the physics in Y$_ 3$Cu$_9$(OH)$_{19}$Cl$_8$, one important perspective to complete our understanding of this material is to quantify and address the role of Dzyaloshinskii-Moriya interaction. Let us indeed recall that this anisotropy of the interaction was pointed out as the main ingredient driving long range order in the sister compound YCu$_3$(OH)$_6$Cl$_3$~\cite{Arh20} with a perfect kagome lattice. Quantum fluctuations and especially their role at the phase boundaries and how they impact the "jammed" classical spin liquid phase is another key perspective of this work. Finally we note that a Br counterpart \cite{zeng2022possible} was recently synthesized, showing no sign of ordering, and likely allowing to span further the phase diagram of the anisotropic kagome model. 

\section*{Acknowledgements}
The authors acknowledge the support of the French Agence Nationale de la
Recherche, under Grant No. ANR-18-CE30-0022 "LINK", and funding from Deutsche Forschungsgemeinschaft
(DFG) through TRR 288-422213477 (project A03). We thank Samir Hammoud for carrying out the ICP and gas extraction experiments, and A. Razpopov, R. Valenti, H.O. Jeschke and J. Reuther for useful discussions and for sharing INS simulation data.

%

\end{document}